\documentclass[a4paper,fleqn,usenatbib]{mnras}

\usepackage{newtxtext,newtxmath}

\usepackage[T1]{fontenc}
\usepackage{ae,aecompl}

\usepackage{graphicx}	
\usepackage{amsmath}	
\usepackage{amssymb}	
\usepackage{times}
\usepackage{graphics, subfigure}
\usepackage{epsfig}
\usepackage{multirow}
\usepackage{ulem}
\usepackage{pdfpages}

\def\simlt{\mathrel{\hbox{\rlap{\hbox{\lower4pt\hbox{$\sim$}}}\hbox{$<$}}}}
\def\simgt{\mathrel{\hbox{\rlap{\hbox{\lower4pt\hbox{$\sim$}}}\hbox{$>$}}}}

\newcommand{\mysim}{\mathord{\sim}}

\newcommand{\myapprox}{\mathord{\approx}}

\title[Numerical calculation of detonation waves]{An accurate and efficient numerical calculation of detonation waves in multidimensional supernova simulations using a burning limiter and adaptive quasi-statistical equilibrium}

\author[D. Kushnir and B. Katz]{
Doron Kushnir$^{1}$\thanks{E-mail: doron.kushnir@weizmann.ac.il} and Boaz Katz$^{1}$
\\
$^{1}$Dept. of Particle Phys. \& Astrophys., Weizmann Institute of
Science, Rehovot 76100, Israel
}

\date{Accepted XXX. Received YYY; in original form ZZZ}

\pubyear{2019}

\begin{document}
\label{firstpage}
\pagerange{\pageref{firstpage}--\pageref{lastpage}}
\maketitle

\begin{abstract}
Resolving the small length-scale of thermonuclear detonation waves (TNDWs) in supernovae is currently not possible in multidimensional full-star simulations. Additionally, multidimensional simulations usually use small, oversimplistic reaction networks and adopt an ad hoc transition criterion to nuclear statistical equilibrium (NSE). The errors due to the applied approximations are not well understood. We present here a new accurate and efficient numerical scheme that accelerates the calculations by orders of magnitudes and allows the structure of TNDWs to be resolved. The numerical scheme has two important ingredients: (1) a burning limiter that broadens the width of the TNDW while accurately preserving its internal structure, and (2) an adaptive separation of isotopes into groups that are in nuclear statistical quasi-equilibrium, which resolves the time-consuming burning calculation of reactions that are nearly balanced out. Burning is calculated in situ employing the required large networks without the use of post-processing or pre-describing the conditions behind the TNDW. In particular, the approach to and deviation from NSE are calculated self-consistently. The scheme can be easily implemented in multidimensional codes. We test our scheme against accurate solutions of the structure of TNDWs and against homogeneous expansion from NSE. We show that with resolutions that are typical for multidimensional full-star simulations, we reproduce the accurate thermodynamic trajectory (density, temperature, etc.) to an accuracy that is better than a percent for the resolved scales (where the burning limiter is not applied), while keeping the error for unresolved scales (broadened by the burning limiter) within a few percent.
\end{abstract}




\section{Introduction}
\label{sec:Introduction}

Thermonuclear detonation waves (TNDWs) are believed to play a key role in supernovae \citep[][]{Hoyle1960,Fowler1964}. The TNDW structure is important for both the energy release and the nucleosynthesis during the explosion, and it is therefore a crucial ingredient for supernova modelling \citep[see][for a recent review]{Seitenzahl2017}. However, resolving the TNDW structure in a multidimensional hydrodynamical simulation of a supernova is currently impossible. This is because the fast thermonuclear burning dictates a burning length-scale that is much smaller than the size of the star, and because the number of isotopes participating in the thermonuclear burning is very large. These problems have led to the introduction of various approximations to multidimensional hydrodynamical simulations of full stars. The error introduced by these approximations is, however, not well understood. 

The various approximations that were introduced to address the small burning scale so far can be roughly divided into two classes. Some codes use an exact solution to a steady-state TNDW, which is incorporated in the time-dependent simulation by imposing the position of the TNDW and/or the thermonuclear burning that follows \citep[e.g.,][]{Golombek2005,Calder2007,Fink2011,Townsley2016,Miles2019}. It is not clear, however, how much the actual flow deviates from the steady-state assumption, and the resulting error is hard to estimate. Moreover, the error introduced by the steady-state assumption cannot be controlled by increasing the resolution. 

Most other codes (in particular, VULCAN \citep{Livne1993IMT} and FLASH \citep{Fryxell2000}) resolve the fast temporal evolution by employing many short burning time-steps within each hydrodynamical step (with constant density and temperature throughout the hydrodynamical step). In principle, with increasing spatial resolution, correct converged results can be obtained. In practice, however, such a high spatial resolution is impractical in global multidimensional simulations, and the size of the cells is much larger than the burning scale. The actual hydrodynamical trajectory within the hydrodynamical step is therefore not accurately captured. Because of the large gap between the achievable size of the cells and the burning scale, it is hard to trust the results, even when they appear to converge. 

In addition to the coarse resolution, most global multidimensional simulations use a short isotope list during the hydrodynamic calculation, with a following post-processing step, in which the burning is recalculated with detailed reaction network for a small subset of trace-particle trajectories.  The convergence and consistency of the obtained results are hard to demonstrate. In some cases, the forward and backward reactions of many isotopes reach near balance, resulting in stiff burning ODEs, which make the calculations of burning even harder. The most extreme case is of nuclear statistical equilibrium (NSE), in which a direct solution of the ODEs becomes impractical. However, in this case the abundances can be calculated algebraically. Severe detailed balance problem may occur in places where full NSE is not reached, and cannot be avoided in this way. 

An excellent test for codes used for global hydrodynamical simulations is a steady-state, planar detonation wave, given by the ZND theory \citep{ZelDovich1940,Doring1943,vonNeumann47}, which can be derived accurately for the case of TNDWs by solving ODEs \citep[][see Appendix~\ref{sec:structure} for details]{Imshennik1984,Khokhlov89,Townsley2016,Kushnir2019}. Only in a few cases such codes were tested against this solution. Moreover, many of the published tests focus on conditions in which the main challenges associated with global simulations are not recovered. In some cases, the resolution is too high to be practical for full-star simulations \citep{Boisseau1996,Noel2007}, and in other cases the upstream density, $\rho_0$, is low, such that the burning length-scale of the TNDW is large \citep[][]{Townsley2012}. A notable exception is the code test that was presented by \citet{Gamezo99}, where an Eulerian code was used to simulate a steady-state, planar TNDW where the burning front is not resolved. While they obtained a good agreement with the accurate solution, the results are not conclusive as they imposed the full accurate solution as their boundary condition, and the numerical solution in the first $\mysim10$ cells following the shock wave is not presented. 

As we describe in Section~\ref{sec:TNDW challenge} the case of TNDWs propagating in a plasma with equal mass fraction of $^{12}$C and $^{16}$O (CO) and density of $\rho_{0,7}\approx1$ \footnote{We define $\rho_{x}\equiv\rho\,[\textrm{g}\,\textrm{cm}^{-3}]/10^{x}$ and $\rho_{0,x}\equiv\rho_{0}\,[\textrm{g}\,\textrm{cm}^{-3}]/10^{x}$.}, which is typical for Type Ia supernovae, is particularly challenging for full-star simulations. In addition to the problem that the burning length-scale is much smaller than the typical cell size, near detailed balance is obtained for many isotopes while NSE is not reached. We test in Section~\ref{sec:TNDW challenge} two available one-dimensional (1D) codes: a modified version of the 1D, Lagrangian version of the VULCAN code \citep[hereafter V1D; for details, see][]{Livne1993IMT} and a modified version of the Eulerian, 1D hydrodynamic FLASH4.0 code with thermonuclear burning \citep[][]{Fryxell2000,dubey2009flash}, against the $\rho_{0,7}=1$ case. We show that with resolutions that are typical for multidimensional full-star simulations, the V1D and the FLASH results are not satisfactory (up to $50\%$ error in V1D and up to $20\%$ error in FLASH). We demonstrate in Section~\ref{sec:demonstration} the operation of a new numerical scheme for thermonuclear burning that can be implemented in multidimensional full-star simulations. The new scheme allows an accurate calculation of TNDWs in a consistent way (i.e., without pre-describing the position and/or the conditions behind the TNDW) with all thermonuclear burning taking place in situ (without post-processing) for an arbitrary reaction network with hundreds of isotopes. The new scheme contains two important ingredients: (1) a burning limiter \citep[a variant of][]{Kushnir2013}, which guarantees that the thermodynamic variables and the composition are accurate for the resolved scales, while keeping the numerical thermodynamic trajectory for unresolved scales within some controlled error from the true thermodynamic trajectory, and (2) adaptive statistical equilibrium (ASE) burning, which groups isotopes that are in detailed balance into one effective isotope, where the ratio between the isotope abundances inside the group is given from equilibrium conditions \citep[this is an extension of the earlier attempts of][]{Hix2007,Parete-Koon2008b,Parete-Koon2008,Parete-Koon2010}. 
  
The main results of the paper are presented in Section~\ref{sec:demonstration}, and the rest of the paper contains a detailed description of the numerical scheme (Section~\ref{sec:new scheme}), along with extensive tests against a steady-state, planar TNDW (Sections~\ref{sec:new scheme} and~\ref{sec:more cases}). We test at a large range of upstream densities, and also with a composition of pure $^{4}$He (He). An additional test of expansion from NSE is presented in Section~\ref{sec:expansion}, where we also present a new accurate ODE solver for this problem. We summarize our results in Section~\ref{sec:discussion}. 

Some aspects of this work were calculated with a modified version of the {\sc MESA} code\footnote{Version r7624; https://sourceforge.net/projects/mesa/files/releases/.} \citep{Paxton2011,Paxton2013,Paxton2015}. 


\section{The challenge of calculating TNDW\lowercase{s}}
\label{sec:TNDW challenge}

\begin{figure}
\includegraphics[width=0.47\textwidth]{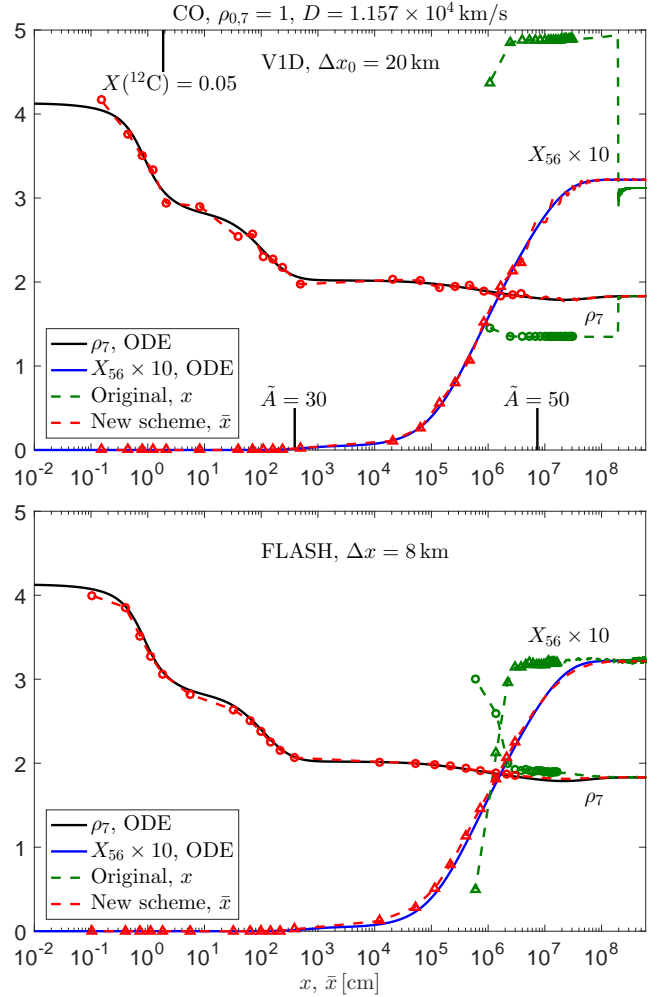}
\caption{The density and $^{56}$Ni mass fraction profiles for a slightly overdriven detonation wave in CO with $\rho_{0,7}=1$ and $D=1.157\times10^{4}\,\textrm{km}\,\textrm{s}^{-1}$, as a function of the distance behind the shock. The ODE solution (black and blue) is compared with the results obtained with V1D (upper panel, initial resolution of $\Delta x_{0}=20\,\textrm{km}$) and FLASH (lower panel, fixed resolution of $\Delta x=8\,\textrm{km}$) after the shock traversed a typical dynamical scale of $v/\sqrt{G\rho_{0}}$ with $v=10^{4}\,\textrm{km}\,\textrm{s}^{-1}$ (green; the first 20 cells are marked with a circle for the density and with a triangle for $X_{56}$). The resolutions of the two codes are comparable, after taking into account the $\mysim2-4$ compression behind the shock. The smallest $x$ value corresponds to the cell immediately behind the shock wave (note that we allow a small shift, up to a few cells, when comparing with the ODE solution, so the size of the cells is most easily read by the distance between the two leftmost points). The large $x$ values are further away from the shock and closer to the piston (V1D) or solid wall (FLASH) boundary condition (see Section~\ref{sec:new scheme} for details). The results in red were obtained with our new scheme, and they are presented as a function of $\bar{x}$ (see text for details) in order to demonstrate that the thermodynamic trajectory is completely resolved. The new scheme density profiles are presented in Figures~\ref{fig:ASE_Demo_CO_1e7} and~\ref{fig:New_scheme_CO_1e7} as a function of $x$. Bars mark the positions where $X(^{12}\rm{C})=0.05$ and $\tilde{A}=30,50$ in the ODE solution. 
\label{fig:Failure2_CO_1e7}}
\end{figure}

The main challenge associated with simulating TNDWs is demonstrated in Figure~\ref{fig:Failure2_CO_1e7}, where the exact structure obtained through the solution of the ODEs is compared to the results obtained by the hydrodynamic simulations V1D and FLASH. The density and the mass fraction of $^{56}$Ni, $X_{56}$, profiles of an unsupported, steady-state, planar TNDW, for $\rho_{0,7}=1$, upstream temperature $T_{0,9}=0.2$ \footnote{We define for the temperature $T_{x}\equiv T\,[\textrm{K}]/10^{x}$ and for the upstream temperature $T_{0,x}\equiv T_0\,[\textrm{K}]/10^{x}$.} (used throughout the paper), and CO is presented, as a function of the distance behind the shock wave, $x$ (solid lines)\footnote{The pathological detonation velocity in this case is $D_{*}=1.1560\times10^{4}\,\textrm{km}\,\textrm{s}^{-1}$, so we use a detonation velocity of $D=1.157\times10^{4}\,\textrm{km}\,\textrm{s}^{-1}$ (\textit{slightly overdriven}).}. The input physics for this calculation, which contains $178$ isotopes and is used throughout the paper, is given in Appendix~\ref{sec:input}. A brief summary of the steady-state, planar TNDW solution is given in Appendix~\ref{sec:structure}. The structure of the same detonation wave can be calculated using hydrodynamical codes, which are used to solve multidimensional, time-dependant supernova explosions. The profiles that were obtained with the two codes, with typical highest resolutions that are currently possible for three-dimensional supernova simulations (the resolutions of the two codes are comparable, after taking into account the $\mysim2-4$ compression behind the shock), are shown for comparison (green dashed lines, circles for the density and triangles for $X_{56}$, a detailed description of the set-up of the calculations is given in Section~\ref{sec:new scheme}). As can be seen, the scales on which the structure evolves and significant amount of $^{56}$Ni is synthesized, $x\lesssim10^{6}\,\textrm{km}$, are not resolved. While for sufficiently large $x$, the numerical solutions approach the correct end (NSE) state, since this is a consequence of a steady-state configuration that conserves energy, there is no guarantee that the actual trajectory on intermediate scales ($x\sim10^{6}-10^{7}\,\textrm{cm}$) would be accurate. Indeed, the solution obtained by V1D misses the correct density by $20-30\%$ and the correct $X_{56}$ by a factor of $\sim2$, and then a weak shock wave raises the density to the correct NSE values. While FLASH performs better here, with a deviation of $\sim5\%$ in density and a deviation of $\sim 20\%$ in $X_{56}$ on intermediate scales \citep[our results are very similar to results of][for their parametrized burning scheme implemented in FLASH, see their figures 9 and 10]{Townsley2016}, significant errors can accumulate in full-star explosion simulation. For example, \citet{Miles2019} found errors of order unity in the total synthesized $^{56}$Ni mass in a central detonation of a $0.8\,M_{\odot}$ WD (with a central density of $\sim10^{7}\,\textrm{g}\,\textrm{cm}^{-3}$) with such resolutions (see their table 1). In addition, even with these low resolutions, it is currently impossible to use the required large number of isotopes in multidimensional simulations. What make matters worse, is that as the material approaches NSE, numerical accuracy is being lost because of the detailed balance of fast reactions, leading to slow convergence and possible errors in the burning iterations. 

\begin{figure}
\includegraphics[width=0.48\textwidth]{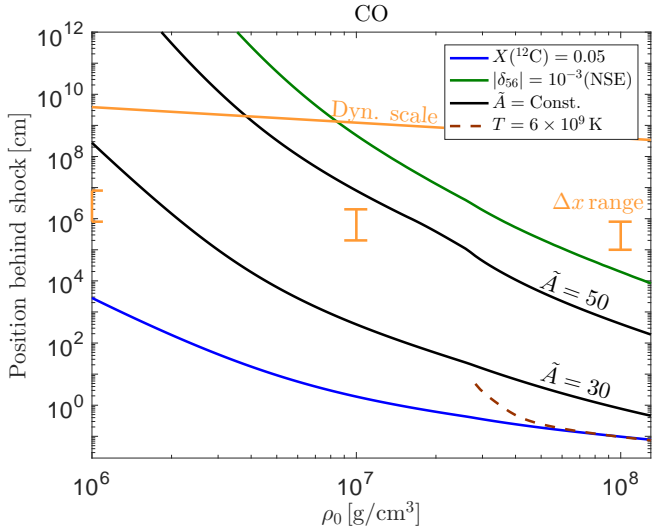}
\caption{Different scales of the CO detonation in comparison with a typical dynamical scale of $v/\sqrt{G\rho_{0}}$ with $v=10^{4}\,\textrm{km}\,\textrm{s}^{-1}$ (orange). Shown are the $^{12}$C consumption scale (blue), the location where $|\delta_{56}\equiv\ln\left(X(^{56}\rm{Ni})/X^{*}(^{56}\rm{Ni})\right)|=10^{-3}$ (where $X^{*}$ is the NSE composition calculated according to Equation~\eqref{eq:NSE2}, such that the material is very close to NSE, green), the location where $\tilde{A}=30$, and $50$ (bottom to top, black), and the location where $T_9=6$ (dashed brown). Also shown are the range of resolutions that are considered in this paper for a few upstream density values. These resolutions are typical for multidimensional hydrodynamical simulations of supernovae, and are $\mysim1000$ times smaller than the size of the star, which we approximate here as the dynamical scale. 
\label{fig:CO_DetonationProp}}
\end{figure}

We can separate the numerical problem into three challenges \citep[see e.g.][]{Townsley2016}:
\begin{enumerate}

\item \textbf{Challenge I: the burning scales are much smaller than the typical cell size.} Following some induction time, the $^{12}$C is consumed and $\myapprox0.3\,\textrm{MeV}/m_{p}$ are released (indicated in Figure~\ref{fig:Failure2_CO_1e7} at $x\approx2\,\textrm{cm}$). This is followed by $^{16}$O burning, in which additional $\myapprox0.3\,\textrm{MeV}/m_{p}$ are released. In order to follow the synthesis of heavy elements, it is useful to define the average nucleon number of the heavy elements
\begin{eqnarray}\label{eq:Ytilde def}
\tilde{A}=\frac{1}{\tilde{Y}}\sum_{i,i\ne n,p,\alpha}Y_{i}A_i,\;\;\tilde{Y}=\sum_{i,i\ne n,p,\alpha}Y_{i},
\end{eqnarray}
where $Y_{i}\approx X_{i}/A_{i}$ are the molar fractions of the nuclei, $X_{i}$ are the mass fractions, and $A_{i}$ are the nucleon numbers. The $^{16}$O burning produces intermediate-mass elements, reaching $\tilde{A}\approx30$ at $x\approx400\,\textrm{cm}$. The next burning stage is heavy elements synthesis, which spans many orders of magnitudes and ends with $\tilde{A}\approx50$ at $x\approx7\times10^{6}\,\textrm{cm}$. As can be seen, both the scales over which most of the energy is being released and most of heavy elements are synthesized are not resolved. This situation becomes worse at higher densities, as shown in Figure~\ref{fig:CO_DetonationProp}, in which these scales are presented for accurate (ODE) solutions performed over a range of upstream densities. As can be seen in the figure, the scales over which the $^{12}$C is consumed and energy is released (blue line), and the scales over which heavy elements are synthesized (black lines) are smaller than the typical resolutions in multidimensional hydrodynamical simulations of supernovae (shown as orange bars). The first challenge is to calculate correctly the state of the material, despite the smallness of the burning scales. 

\item \textbf{Challenge II: a prohibitive large number of isotopes are required for an accurate calculation.} Currently, the most sophisticated global three-dimensional simulations use no more than $\mysim10$ isotopes, while more than $\mysim150$ isotopes are required to reach $1\%$ level accuracy \citep[][see figures 13 and 22 of that paper]{Kushnir2019}. In particular, commonly used $\alpha$-networks lead to significant errors, due to missing important non-$\alpha$-isotopes \citep{Kushnir2019}. Using long isotope lists is challenging both in terms of computation time (naively scales as the number of the isotopes square) and of memory requirement. The second challenge is therefore to calculate the thermonuclear burning in an efficient way despite the large number of required isotopes. 

\item \textbf{Challenge III: detailed balance of fast reactions leads to slow convergence and possible errors in the burning iterations.} The fast forward and backward reactions of many isotopes reach near balance, resulting in stiff ODEs, challenging the burning calculation. For example, the reaction $^{28}$Si$(\alpha,p)^{31}$P in the solution presented in Figure~\ref{fig:Failure2_CO_1e7}, has a high forward rate, $\dot{Y}_f$, with a time-scale $t_{28}\equiv Y(^{28}$Si$)/\dot{Y}_f$ that is smaller by a factor of more than $10^{6}$ than the typical sound crossing time of the numerical cell at $x\sim10^{7}\,\textrm{cm}$. The burning calculation becomes particularly challenging, when the plasma approaches NSE and most reactions involve time-scales much smaller than the sound crossing time. The location at which the material approached NSE is shown in Figure~\ref{fig:CO_DetonationProp} for a wide range of upstream densities (green line, defined as the position where $X(^{56}$Ni$)$ approaches its NSE value within $10^{-3}$, see detailed discussion in Section~\ref{sec:eq burning}). As can be seen, for $\rho_{0,7}\gtrsim1$ the material is close to NSE at scales that are smaller than the dynamical scale (orange line), and therefore NSE is reached in explosions involving such densities, challenging their simulation. 

A note is in place regarding the common practice to resolve this challenge, by choosing a threshold temperature, $T_{\textrm{NSE}}$, above which the material is assumed to be in NSE \citep[see detailed discussion in][]{Paxton2015}. For such cells, the composition is then not determined from evolving the rate equations, but rather from the equilibrium condition. While this approach is useful, it is not clear if a threshold temperature can correctly capture the transition to NSE. For example, the position of $T=6\times10^{9}\,\textrm{K}$ is shown with brown dashed line in Figure~\ref{fig:CO_DetonationProp}, and does not coincide with the location of transition to NSE (green line). In fact, for upstream densities $\rho_{0,7}\lesssim3$ the material never reaches this temperature, even though NSE is approached at scales which are orders of magnitude smaller than the dynamical scale (this problem will become worse for higher threshold temperatures), while for upstream densities $\rho_{0,7}\gtrsim3$ the material reaches this temperature on scales in which the material is far from NSE (this problem will become worse for lower values of threshold temperatures)\footnote{\citet{Bravo2019} suggested to use a density threshold of $8\times10^{7}\,\textrm{g}\,\textrm{cm}^{-3}$ for imposing the NSE condition, which has similar drawbacks.}. 

\end{enumerate}

\section{A new burning scheme}
\label{sec:demonstration}

The new burning scheme contains two important ingredients, both following the principle that large changes of the variables in a cell are not allowed during a sound crossing time of the cell, $t_{s}\equiv\Delta x/c_{s}$, where $\Delta x$ is the size of the cell and $c_s$ is the speed of sound. The first ingredient is a burning limiter (Section~\ref{sec:limiter demo}), which addresses the first challenge in Section~\ref{sec:TNDW challenge}, that the burning scales are much smaller than the typical cell size. The burning limiter multiplies all reaction rates by the same factor, $f_{lim}$, stretching the burning front over many cells \citep[for a different broadening method, see][and discussion in Section~\ref{sec:others limiter}]{Shen2018}, and allowing it to be resolved without changing the thermodynamic trajectory. The second ingredient is ASE burning (Section~\ref{sec:ASE}), which addresses the challenge that the material can approach NSE over scales much smaller than the dynamical scale and partially resolve the challenge that a large number of isotopes are required for an accurate calculation (we are providing a solution for the computation time but not for the memory issue). This is achieved by adaptively grouping isotopes that are in detailed balance into effective isotopes. The evolution of the effective isotopes is calculated by solving the effective rate equations ODEs, while the individual isotopes within the groups are found algebraically by solving generalized equations of nuclear statistical quasi-equilibrium \citep[NSQE;][]{Bodansky1968}. In this way the computation time is significantly reduced, while preserving the accurate evolution of the original large number of isotopes, and allowing a self-consistent transition to NSE regimes. Here, we describe the principles of the new scheme and demonstrate its performance, as implemented in V1D and FLASH, against the same TNDW solution described in Section~\ref{sec:TNDW challenge}. A detailed description of the prescriptions is given in Section~\ref{sec:new scheme}. 

\subsection{A burning limiter -- to resolve small scales}
\label{sec:limiter demo}

We use a variant of the burning limiter that was suggested by \citet{Kushnir2013} to prevent unstable numerical burning. All reaction rates are multiplied by an adaptive factor, $f_{lim}$, set by combination of criteria that limit the changes of energy and composition due to thermonuclear burning within a sound crossing time of the numerical cell. The outcome of the limiter application, is a widening of the burning front across a (fixed) large number of cells. With higher resolutions the widened burning front becomes smaller, and convergence is achieved once the widened front is much smaller than the dynamical scales of the system. In this situation, the widened front is in an approximate steady state in the shock frame. The thermodynamic trajectory of fluid elements across the steady-state region is independent of $f_{lim}$, and is calculated accurately, without the need for extreme resolution. The only effect of $f_{lim}$ is to slow down the reaction rates, and any thermodynamic state of a given fluid element in real time, $\bar{t}$, is shifted to a simulation time, $t$, related by
\begin{eqnarray}\label{eq:tbar}
t=\int_{0}^{\bar{t}}\frac{1}{f_{lim}(t')}dt',
\end{eqnarray}
where the integral is over the time since the fluid element crossed the shock (see discussion in Section~\ref{sec:limiter}). Similarly, we can relate the real distance behind the shock, $\bar{x}$ to the simulation distance, $x$. Unless stated otherwise, the burning limiter that we use limits the changes in both energy and composition to a fraction $f=0.05$ during a sound crossing time (see details in Section~\ref{sec:limiter}).  

The density and the $X_{56}$ profiles that were obtained with the burning limiter implemented in V1D and FLASH, for the same case discussed in Section~\ref{sec:TNDW challenge}, are compared to the ODE solution in Figure~\ref{fig:Failure2_CO_1e7} (red lines). The profiles are presented as a function of $\bar{x}$  (where $f_{lim}(t)$ used in the coordinate transformation Equation~\eqref{eq:tbar}, is calculated based on the conditions obtained by the ODE) in order to allow direct comparison to the entire thermodynamic trajectory. As can be seen, good agreement is obtained throughout the burning front (errors of a few percents). The region in which the limiter is operating (such that $\bar{x}$ is significantly smaller than $x$) contains $\mysim1/f=20$ cells. Following this region the simulation distance quickly approaches the true distance ($x\approx\bar{x}$), and the error of the calculated profiles is on the sub-percent level. This demonstrates that the use of a burning limiter allows the thermodynamic history of any fluid element to be accurately calculated without the use of an extreme resolution. The same density profiles are presented in Figures~\ref{fig:ASE_Demo_CO_1e7} and~\ref{fig:New_scheme_CO_1e7} as a function of $x$.

\subsection{Adaptive statistical equilibrium (ASE) -- to allow consistent approach to NSE and large networks}
\label{sec:ASE}

\begin{figure}
\includegraphics[width=0.48\textwidth]{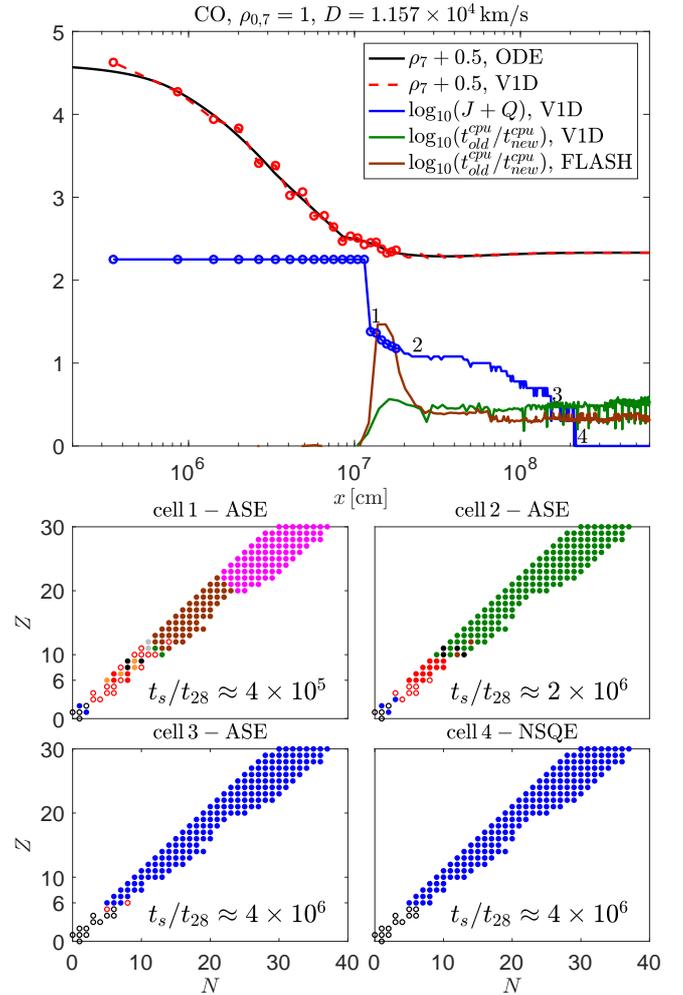}
\caption{The operation of the ASE scheme. Red: the density profiles for a slightly overdriven detonation wave in CO with $\rho_{0,7}=1$ and $D=1.157\times10^{4}\,\textrm{km}\,\textrm{s}^{-1}$, as a function of the distance behind the shock, obtained with V1D ($\Delta x_{0}=20\,\textrm{km}$) after the shock traversed a dynamical scale. Note that the profiles are plotted as a function of the simulated $x$-coordinate. Black: The ODE solution for the corresponding TNDW (transformed to the $x$-coordinate). Blue: The number of variables for the net integration, $J+Q$, as determined by the ASE scheme (when ASE is triggered there is an additional group that contains $n$, $p$, $\alpha$ and possibly other light isotopes, such that the total number of groups is $J+Q+1$, see text). For small $x$, the reactions are not fast enough to reach detailed balance, so the ASE scheme is not triggered (such that $J=178$, $Q=0$). Slightly after $x=100\,\textrm{km}$ (cell $1$), the reactions are fast enough to form an ASE configuration with $Q=8$ and $J=16$. As the distance behind the shock wave increases, configurations with lower number of groups are formed (cells $2$ and $3$), until a configuration with $J=0$, $Q=1$ (NSQE, e.g., cell $4$) is formed. The configurations in cells $1-4$ are presented in the lower panels. Each isotope is a black circle (if it is included in the light isotope group), a red circle (if it is not grouped), or is coloured according to its group. We provide for each configuration the value of $t_s/t_{28}$, which is discussed in the text. As the number of groups decreases, the computation time drops. This is demonstrated by the comparison between the cpu time for calculating the burning step with the ASE scheme, $t_{new}^{cpu}$, to the cpu time for calculating the same burning with the original schemes $t_{old}^{cpu}$, both for V1D (green) and for FLASH (brown). We average the ratio $t_{old}^{cpu}/t_{new}^{cpu}$ over two cells to decrease fluctuations in the cpu time. 
\label{fig:ASE_Demo_CO_1e7}}
\end{figure}

ASE is applied whenever a fast reaction cycle that exchange $\alpha$ with two $n$ and two $p$ can reach detailed balance within a predetermined fraction, $\epsilon$, of the sound crossing time, $t_s$. When ASE is applied, the other isotopes are grouped into the effective isotopes, based on the reaction rates. Any two isotopes that can be converted by a fast reaction with a time-scale less than $\epsilon t_s$ are grouped together (see detailed discussion in Section~\ref{sec:eq burning}). An effective isotope that include $n$, $p$, and $\alpha$ (potentiality with additional isotopes) is always included. The operation of the ASE scheme is demonstrated in Figure~\ref{fig:ASE_Demo_CO_1e7} for the case discussed above (note that this time the profiles are plotted as a function of the simulated $x$). Shown in blue is the total number, $J+Q$, of isolated isotopes, $J$, and of effective isotopes, $Q$ (when ASE is triggered there is an additional group that contains $n$, $p$, $\alpha$ and possibly other light isotopes, such that the total number of groups is $J+Q+1$). For small $x$, the reactions are not fast enough to reach detailed balance, so the ASE scheme is not triggered (such that $J=178$, $Q=0$). Slightly after $x=100\,\textrm{km}$ (cell $1$ in the figure), the reactions are fast enough to form an ASE configuration with $Q=8$ and $J=16$. This configuration (and other configurations) are presented in the lower panels of the figure. As the distance behind the shock wave increases, configurations with lower number of groups are formed (cells $2$ and $3$), until a configuration with $J=0$, $Q=1$ (NSQE, e.g., cell $4$) is formed. The reaction rates between this one group and the group of light elements are slow enough to be resolved during a cell crossing time, and a state of NSE is not reached in this example. Provided for each configuration in the appropriate lower panel, are the values of $t_s/t_{28}$, the ratio between the cell sound crossing time, and the time-scale of the forward rate of the reaction $^{28}$Si$(\alpha,p)^{31}$P, discussed in Section~\ref{sec:TNDW challenge}. This ratio reaches $\mysim10^{6}$, and solving separately for $Y_{28}$ and $Y_{31}$ is inefficient. The ratios between the cpu time for calculating the burning step with the ASE scheme to the cpu time for calculating the same burning with the original V1D (green) and FLASH (brown) are shown. The original V1D and FLASH burning scheme are slower by roughly a factor of $3$ (note that for a few cells the original FLASH burning scheme is slower by factor $\mysim30$, which has a large effect on the total runtime of the simulations, see examples in Section~\ref{sec:new scheme}).

\section{A detailed description of the new burning scheme}
\label{sec:new scheme}

In this section, we present in detail our new numerical scheme for thermonuclear burning. We first describe the set-up and the current coupling of burning to the hydrodynamics in V1D (Section~\ref{sec:old scheme V1D}) and in FLASH (Section~\ref{sec:old scheme FLASH}) that are used to simulate TNDWs without the implementation of the new scheme. The two new ingredients of our scheme are described in Section~\ref{sec:limiter} (a burning limiter) and in Section~\ref{sec:eq burning} (ASE). We study the convergence properties of the new scheme in Section~\ref{sec:convergence} and we provide a few example calculations in Section~\ref{sec:example results}.

\subsection{Lagrangian code -- VULCAN}
\label{sec:old scheme V1D}

We modified V1D to be compatible with the input physics of Appendix~\ref{sec:input}. We do not use linear artificial viscosity, the Courant time-step factor is $0.25$, and the maximum relative change of the density in each cell during a time-step is set to $0.01$. Burning is not allowed on shocks (identified as cells where $q_{v}/p>0.1$, where $q_{v}$ is the artificial viscosity and $p$ is the pressure). The allowed error tolerance for the burning integration is $\delta_{B}=10^{-8}$ (see below). Initially, all cells have equal size, $\Delta x_{0}$. 

We perform the simulation in the downstream frame of the detonation wave with the downstream on the left-hand side and the upstream on the far right-hand side. Fluid velocities are often described in the shock frame and are then denoted by $u$. A calculated (overdriven) detonation wave has a detonation velocity $D$ (compared to the upstream) and a final NSE state at $x=\infty$, marked with a subscript $*$ with a velocity $u_*$ (compared to the upstream). In the (simulated) downstream frame, the incoming upstream has a velocity $u_{*}-D$, and is propagating towards the left-hand boundary. 

We implement a piston boundary condition at the left-hand boundary, with the purpose of converging quickly to the steady-state solution. At the NSE state, the fluid is at rest in the downstream frame, and the velocity of the piston is zero. At earlier times, the fluid at the left-hand boundary has a non-zero velocity $u(t)-u_{*}$, which is known from the ODE calculations and is imposed as the velocity of the piston. Time is chosen such that $t=0$ is the time when the shock crossed the left-hand boundary. In practice, the ODE solution is used only up to some finite time $t_{f}$ (because of accuracy issues, see detailed discussion in Section~\ref{sec:expansion}), beyond which the velocity is smoothly reduced to zero exponentially. In all cases $u(t_f)-u_{*}$ is smaller than $1\,\textrm{km}\,\textrm{s}^{-1}$, and the artificial piston velocity at $t>t_{f}$ has a negligible effect on our results. We also use a simpler boundary condition where the velocity of the piston is set to zero at all time. While less accurate, this boundary condition can be easily implemented in Eulerian codes, allowing direct comparison to Lagrangian codes. The size of the computed domain is chosen such that the shock wave reaches the right-hand boundary (that is moving with $u_{*}-D$ at all times) roughly after a gravitational dynamical time-scale for the simulated upstream density.

In the beginning of each hydrodynamical step, the density, $\rho^n$, temperature, $T^n$, and composition, $X_{i}^{n}$, of a given numerical cell are known from the previous calculations. Following the force calculation over the hydrodynamical time-step, $\Delta t$, the density at the end of the time-step, $\rho^{n+1}$, is found. The task of the burning scheme is to solve for the temperature and the composition at the end of the time-step. First, the energy release from burning is calculated with a predictor--corrector scheme. A burning step over $\Delta t$ is calculated with the initial $\rho^n$ and $T^n$ to get an estimate for the thermonuclear energy release, $\Delta q$, and the composition at the end of the time-step $X_{i}^{n+1}$. An estimate for the internal energy per unit mass at the end of the time-step is calculated as
\begin{equation}\label{eq:predictor}
\varepsilon^{n+1}=\varepsilon^{n}-\left(\frac{1}{\rho^{n+1}}-\frac{1}{\rho^n}\right)(p^n+q_{v}^n)+\Delta q,
\end{equation}
where $p^{n}$ and $q_v^n$ are the pressure and the artificial viscosity at the beginning of the time-step. We then have an estimate for the temperature at the end of the time-step, $T^{n+1}=T(\rho^{n+1},\varepsilon^{n+1},X_{i}^{n+1})$. A corrector step follows to better estimate $\Delta q$ and $X_{i}^{n+1}$, which begins with a burning step with $\rho^{n+1}$, $T^{n+1}$, and $X_{i}^{n}$ (note that we use the density and the temperature at the end of the time-step, for a reason that will be clarified in Section~\ref{sec:eq burning}), followed by the solution of
\begin{eqnarray}\label{eq:corrector}
&&\varepsilon^{n+1}=\varepsilon^{n}- \\\nonumber 
&&\left(\frac{1}{\rho^{n+1}}-\frac{1}{\rho^n}\right)\left(\frac{p^{n}+p^{n+1}(\rho^{n+1},\varepsilon^{n+1},X_{i}^{n+1})}{2}+q_{v}^n\right)+\Delta q,
\end{eqnarray}
by iterating $\varepsilon^{n+1}$. Then, we finally have $T^{n+1}=T(\rho^{n+1},\varepsilon^{n+1},X_{i}^{n+1})$. 

The burning step evolves the composition over a time-step $\Delta t$, with constant density, $\rho$, and temperature $T$. The difference in abundances between the beginning and the end of the time-step is used to calculate the energy release
\begin{equation}\label{eq:delta q}
\Delta q=N_{A}\sum_{i}Q_{i}\left(Y_{i}^{n+1}-Y_{i}^{n}\right),
\end{equation}
where $Q_{i}$ are the binding energies of the nuclei. A simple semi-implicit Euler solver with adaptive time-steps is used for the integration. We choose a burning time-step, $\Delta t_{B}=\Delta t$, and iterate with the convergence criterion $\max_{i}(\Delta X_{i})<\delta_B$, where $\Delta X_{i}$ is the change in the composition $X_{i}$ over the last iteration. If the iterations do not converge, then the burning time-step is decreased and another attempt at a solution is carried out. Following a successful iteration procedure, the burning time-step is increased. The process ends when integration along the full $\Delta t$ has been completed. 

\subsection{Eulerian code -- FLASH}
\label{sec:old scheme FLASH}

We modified FLASH to be compatible with the input physics of Appendix~\ref{sec:input}. Instead of using the supplied burning routines of FLASH, which only support hard-wired $\alpha$-nets, we use the burning routines of V1D, including the same integration method. Specifically, instead of using one of the two integration methods supplied with FLASH (either fourth-order Rosenbrock method or variable order Bader--Deuflhard method), we use the much simpler integration scheme of V1D. The main reason for changing the integration method is the easy implementation of our new burning scheme. We find no significant difference between the simple V1D integration scheme and the fourth-order Rosenbrock method in a few cases in which we compare the two (see below). 

The simulations are performed in planar geometry, the cut-off value for composition mass fraction is $\textsc{smallx}=10^{-25}$, and the Courant time-step factor is $\textsc{CFL}=0.2$. Burning is not allowed on shocks and the nuclear burning time-step factor is $\textsc{enucDtFactor}=0.2$. All cells have an equal size, $\Delta x$, which remains constant throughout the simulation (we do not use adaptive mesh refinement in this work). This allows an easy interpretation of our results, with the price of a longer simulation time which is acceptable for our 1D simulations. The simulations are performed in the downstream frame of the detonation wave, and initially the fluid throughout the domain has a velocity $u_{*}-D$. The boundary condition for the left-hand boundary is "reflected" (a solid wall), which allows a direct comparison to Lagrangian codes (see discussion in Section~\ref{sec:old scheme V1D}). We override in each time-step any deviations from the initial conditions of unshocked cells because of the waves that develop next to the right-hand boundary. In this way, cells always have the initial upstream conditions up to the point where the shock crosses them. This can be enforced up to the time when the shock is a few cells away from the right-hand boundary, which is the time that we stop the simulation. The size of the computed domain is chosen such that the shock wave reaches the right-hand boundary roughly after a gravitational dynamical time-scale for the simulated upstream density.

In FLASH, following a hydrodynamical step, we have for each cell $\rho^{n+1}$, $\bar{\varepsilon}^{n+1}$ and $\bar{X}_{i}^{n}$, where $\bar{\varepsilon}^{n+1}$ does not include contribution from burning and $\bar{X}_{i}^{n}$ is different form $X_{i}^{n}$ only because of advection between cells. Then a burning step over $\Delta t$ is calculated with $\rho^{n+1}$ and $\bar{T}^{n+1}=T(\rho^{n+1},\bar{\varepsilon}^{n+1},\bar{X}_{i}^{n})$ to get $\Delta q$ and the compositions at the end of the time-step $X_{i}^{n+1}$. This is followed by an energy update $\varepsilon^{n+1}=\bar{\varepsilon}^{n+1}+\Delta q$, and a temperature update $T^{n+1}=T(\rho^{n+1},\varepsilon^{n+1},X_{i}^{n+1})$. 

\subsection{A burning limiter}
\label{sec:limiter}

The motivation for using a burning limiter was discussed in Section~\ref{sec:demonstration}. In principle, we would like to make sure that all important quantities do not change faster than the cell's sound crossing time \citep[and not just the energy, as in the original burning limiter of ][]{Kushnir2013}. After some experimenting, we have come out with the following limiters:
\begin{eqnarray}\label{eq:limiters}
\textrm{Energy limiter}&:&\,f_\varepsilon=\min\left(\frac{\varepsilon}{\left|\dot{q}\right|}\frac{f}{t_s},1\right) \nonumber \\
np\alpha\,\textrm{limiter}&:&\,f_{np\alpha}=\min\left(\frac{Y_{np\alpha}}{\left|\dot{Y_{np\alpha}}\right|}\frac{f}{t_s},1\right) \nonumber \\
\tilde{Y}\,\textrm{limiter}&:&\,f_{\tilde{Y}}=\min\left(\frac{\tilde{Y}}{\left|\dot{\tilde{Y}}\right|}\frac{f}{t_s},1\right)\nonumber\\
\textrm{Full limiter}&:&\,f_{lim}=\min\left[f_\varepsilon,\left(\frac{\tilde{X}}{f_{\tilde{Y}}}+\frac{X_{np\alpha}}{f_{np\alpha}}\right)^{-1}\right],
\end{eqnarray}
where
\begin{eqnarray}\label{eq:Ynpa def}
X_{np\alpha}=\sum_{i=n,p,\alpha}X_{i},\;\;Y_{np\alpha}=\sum_{i=n,p,\alpha}Y_{i}.
\end{eqnarray}
In this choice, we limit the change in composition by grouping all isotopes in two groups, $np\alpha$ and all the rest, and we limit the change in each of them with $f_{npa}$ and $f_{\tilde{Y}}$, respectively. Since we do not want to apply a strong limit for very small abundances, we geometrically average the two composition limiters, according to their mass fractions. We finally take the minimum between the energy limiter and the composition limiter.

\begin{figure}
\includegraphics[width=0.48\textwidth]{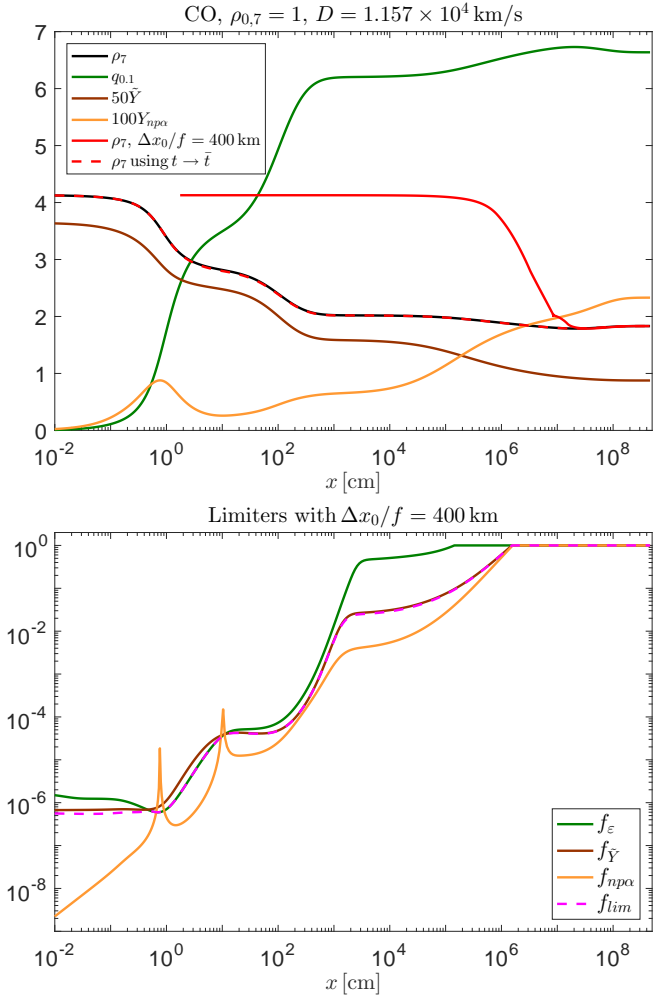}
\caption{The operation of a burning limiter for a slightly overdriven detonation wave in CO with $\rho_{0,7}=1$ and $D=1.157\times10^{4}\,\textrm{km}\,\textrm{s}^{-1}$. Upper panel: profiles from the ODE solution, density (black), thermonuclear energy release ($q_{0.1}=q[0.1\,\textrm{MeV}/m_{p}]$, green), $\tilde{Y}$ (brown), and $Y_{np\alpha}$ (orange), as a function of the distance behind the shock. The density from an ODE solution of the same case with a burning limiter under the assumption of a Lagrangian simulation with $\Delta x_{0}/f=400\,\textrm{km}$ is plotted in red. The same density profile plotted as a function of the original distance behind the shock, by using the transformation of Equation~\eqref{eq:tbar}, is plotted in dashed red. Bottom panel: Profiles of different limiters (see Equations~\eqref{eq:limiters}) for the same case, $f_{\varepsilon}$ (green), $f_{\tilde{Y}}$ (brown), $f_{np\alpha}$ (orange), and $f_{lim}$ (magenta), under the assumption of a Lagrangian simulation with $\Delta x_{0}/f=400\,\textrm{km}$. 
\label{fig:Limiter_Explained_CO_1e7}}
\end{figure}

The operation of the different limiters is demonstrated in Figure~\ref{fig:Limiter_Explained_CO_1e7} for a CO detonation with $\rho_{0,7}=1$ and $D=1.157\times10^{4}\,\textrm{km}\,\textrm{s}^{-1}$. The upper panel shows $\rho$, $q$, $\tilde{Y}$, and $Y_{np\alpha}$ profiles, and the bottom panel shows the different limiters, with the assumption of a Lagrangian simulation with $\Delta x_{0}/f=400\,\textrm{km}$ (note that all limiters depend on $f$ and on the cell size only through $f/t_{s}$, and for Lagrangian simulations the cell size is related to the initial cell size through $\rho\Delta x=\rho_0\Delta x_{0}$). 

First, let us examine the energy limiter behaviour. At a small distance behind the shock wave, the energy release rate is much faster than the sound crossing time, and so a small energy limiter, $f_{\varepsilon}\sim10^{-6}$ is obtained. As we get further away from the shock wave, the energy release rate becomes smaller, and the energy limiter increases, up to a point, $x\sim10^{5}\,\textrm{cm}$, where the energy release time-scale is comparable to the crossing time. Although the energy release time-scale becomes long enough, large variations in composition are possible, as evident from inspection of the $\tilde{Y}$ and $Y_{np\alpha}$ profiles. Indeed, for $x=10^{5}\,\textrm{cm}$, the obtained composition limiters are $<0.1$. In this specific case $X_{np\alpha}$ is quite small, such that the composition limiter is dominated by changes in $\tilde{Y}$. The full limiter follows $f_{\varepsilon}$ and $f_{\tilde{Y}}$ at small distances behind the shock wave and is dominated by $f_{\tilde{Y}}$ at large distances, up to $x\sim10^{6}\,\textrm{cm}$, where the time-scales of all changes are smaller than the sound crossing time. The structure of the detonation wave can be calculated (by solving an ODE) with the application of the burning limiter, and the obtained density is shown in the upper panel. The detonation wave is widened to scales $\lesssim\Delta x_{0}/f$ (note that because of the shock compression $\Delta x<\Delta x_{0}$), and follows the original structure for larger scales. In the upper panel, we plot the burning-limited density profile as a function of the original distance behind the shock, by using the transformation of Equation~\eqref{eq:tbar}. The transformed burning-limited density profile agrees with the exact solution of the ODE, demonstrating that the entire trajectory is reproduced within the widened detonation wave, a fact that cannot be realized by looking at the unresolved plot of density as a function of $x$.

An important property of the burning limiter for a steady-state situation is demonstrated by the form of Equations~\eqref{eq:ZND}. Following a change in composition $dX_{i}$ (that determines the nuclear energy release $dq$), the changes in $d\rho$ and $dT$ are independent of the rate in which this change took place. It follows that if all reaction rates are slower by the same factor, then the fluid reaches the exact same state but over a time longer by the same factor. Therefore, since the burning limiter multiplies all reaction rates by the same factor, $f_{lim}$, it accurately describes the thermodynamic trajectory. This is applicable also to time-dependent TNDWs as long as the length-scale on which the limiter operates is much smaller than any other length-scale of the problem, such that the assumption of a steady state in the region where the limiter operates is justified (see further discussion in Section~\ref{sec:discussion}). 

\begin{figure}
\includegraphics[width=0.48\textwidth]{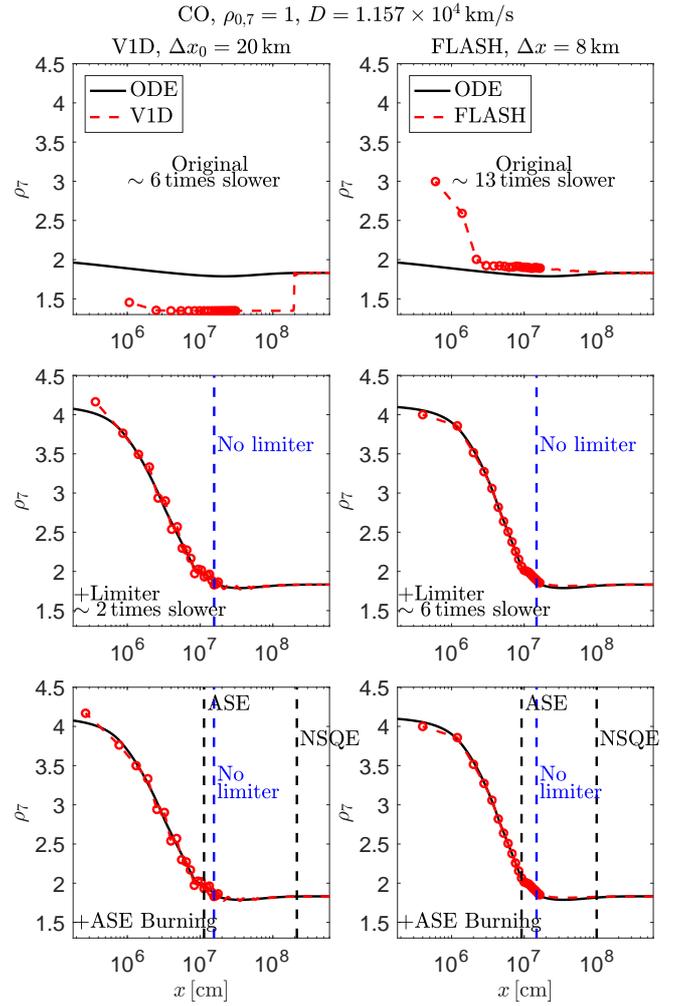}
\caption{The density profiles for a slightly overdriven detonation wave in CO with $\rho_{0,7}=1$ and $D=1.157\times10^{4}\,\textrm{km}\,\textrm{s}^{-1}$, as a function of the distance behind the shock. The ODE solution (black) is compared with the results obtained with V1D (left, $\Delta x_{0}=20\,\textrm{km}$) and FLASH (right, $\Delta x=8\,\textrm{km}$) after the shock traversed a dynamical scale (red, the first 20 cells are marked with a circle). The results at the upper panels were obtained without applying our new burning scheme. The results at middle panels were obtained with the burning limiter and $f=0.05$ (without using ASE). The results at lower panels were obtained with our new scheme (including ASE). Unlike in Figure ~\ref{fig:Failure2_CO_1e7}, the solutions using V1D and FLASH in the middle and bottom panels are shown in the simulation coordinate $x$ rather than the physical coordinate $\bar x$ (see Equation~\eqref{eq:tbar} and discussion that follows). The ODE solution is shown using the corresponding stretched coordinate to allow comparison. The dashed vertical blue line marks the position where the burning limiter is no longer operating in the hydrocodes. The two dashed vertical black lines mark the positions where we first use the ASE scheme and the position where the material has reached NSQE state. The performance speed of the codes (compared to the new burning scheme) is indicated as well.
\label{fig:New_scheme_CO_1e7}}
\end{figure}

The simulated TNDWs in V1D and in FLASH for $\rho_{0,7}=1$, with and without the burning limiter\footnote{Note that the piston boundary velocity in the V1D simulation depends on the ODE solution, which is affected by the burning limiter. The boundary condition used the same burning limiter as in the simulation.} (taking $f=0.05$), were presented in Section~\ref{sec:demonstration}, and they are reproduced in the top and middle panels of Figure~\ref{fig:New_scheme_CO_1e7} for clarity. Note that unlike in Figure~\ref{fig:Failure2_CO_1e7}, the solutions using V1D and FLASH in the middle panel of Figure~\ref{fig:New_scheme_CO_1e7} are shown as a function of the simulation coordinate $x$ rather than the physical coordinate $\bar x$ (see Equation~\eqref{eq:tbar} and discussion that follows). To allow comparison to the ODE solution, it is shown as a function of a corresponding stretched coordinate. We use initial cell sizes of $\Delta x_{0}=20\,\textrm{km}$ for V1D and  $\Delta x=8\,\textrm{km}$ for FLASH (these are comparable, after taking into account the $\mysim2-4$ compression behind the shock), such that most of the nuclear synthesis and the energy release are at scales much smaller than the size of the cells. For the cases without the burning limiter included, we verified that integrating the abundances with the fourth-order Rosenbrock method\footnote{For this we used MESA with option \textsc{rodas4\_solver} and the parameters $\textsc{rtol}=10^{-8}$ (relative error tolerance) and $\textsc{atol}=10^{-9}$ (absolute error tolerance).}, instead of the simple V1D integration scheme, made no effect to our results. The dashed vertical blue line marks the position where the burning limiter is no longer operating in the hydrocodes. 

Consider next the case of CO with $\rho_{0,7}=10$, which is at the high end of detonation upstream densities applicable to the bulk of Type Ia supernova explosions. For this case, $D_{*}=1.1490\times10^{4}\,\textrm{km}\,\textrm{s}^{-1}$, so we use $D=1.150\times10^{4}\,\textrm{km}\,\textrm{s}^{-1}$. We use initial cell sizes of $\Delta x_{0}=8\,\textrm{km}$ for V1D and  $\Delta x=4\,\textrm{km}$ for FLASH, such that the approach to NSE occurs at scales much smaller than the size of the cells (and on scales that are much smaller than the dynamical scale), see Figure~\ref{fig:CO_DetonationProp}. We were unable to calculate this case with the original codes without enforcing NSE conditions, and both codes collapsed after the shock propagated only a few cells. The same results were obtained with the fourth-order Rosenbrock method for integrating the abundances. \citet{Paxton2015} suggested that using extended precision for the integration of the abundances may allow an integration without the need for enforcing an NSE condition. We implemented this suggestion with the recommendation of \citet{Paxton2015}\footnote{Converting the derivative terms in the order they appear from IEEE 64-bit to IEEE 128-bit using the Fortran promotion rules and adding the terms in IEEE 128-bit arithmetic.}, but still the codes collapsed after the shock propagated only a few cells. We therefore enforced $T_{\textrm{NSE}}=6\times10^{9}\,\textrm{K}$ in both codes (detailed description of the NSE enforcement is given in Section~\ref{sec:eq burning}). 

\begin{figure}
\includegraphics[width=0.48\textwidth]{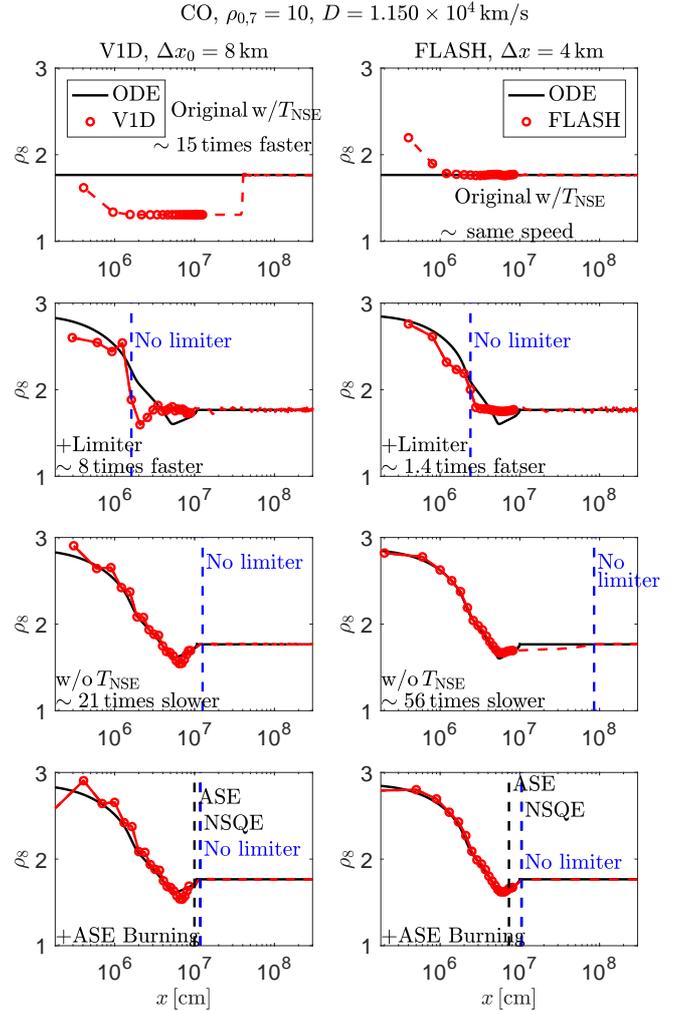}
\caption{Same as Figure~\ref{fig:New_scheme_CO_1e7} for $\rho_{0,7}=10$ and $D=1.150\times10^{4}\,\textrm{km}\,\textrm{s}^{-1}$, $\Delta x_{0}=8\,\textrm{km}$, and $\Delta x=4\,\textrm{km}$. We enforced $T_{\textrm{NSE}}=6\times10^{9}\,\textrm{K}$ for the calculations without the new burning scheme (upper panels). The results at the second (third) row were obtained with the burning limiter and $f=0.05$, with (without) enforcing $T_{\textrm{NSE}}=6\times10^{9}\,\textrm{K}$, and without using ASE. The results at lower panels were obtained with our new scheme (without enforcing $T_{\textrm{NSE}}=6\times10^{9}\,\textrm{K}$ and including ASE). In the lower panels, one dashed black line is underneath the blue dashed line. 
\label{fig:New_scheme_CO_1e8}}
\end{figure}

The density obtained with the original two codes, after the shock traversed a dynamical scale, are compared to the ODE solution of the corresponding detonation wave at the upper panels of Figure~\ref{fig:New_scheme_CO_1e8}. The solution obtained by V1D misses the correct density by $\mysim25\%$, and then a weak shock wave raises the density to the correct NSE values. FLASH performs much better, with less than a percent error. We verified that integrating the abundances with the fourth-order Rosenbrock method, instead of the simple V1D integration scheme, made no effect on our results. 

Calculations with a burning limiter of $f=0.05$ were preformed for this case using the V1D and FLASH simulations (second row of Figure~\ref{fig:New_scheme_CO_1e8}). The $\rho_{0,7}=10$ case is more challenging than the $\rho_{0,7}=1$ case. In this case the limiter stops operating too early, because NSE is imposed once the temperature reaches $T_{\textrm{NSE}}=6\times10^{9}\,\textrm{K}$ (and then reaction rates are not calculated). The use of a burning limiter allows to remove the $T_{\textrm{NSE}}$ constraints, and the results in this case are shown in the third row of Figure~\ref{fig:New_scheme_CO_1e8}. The results from both codes are reasonable, although the limiter is operating for too long in the FLASH code. This is related to the attempt of integrating the rate equations for a material that is very close to NSE, which is also highly inefficient in terms of speed performance. These issues are resolved using the ASE scheme as described in Section~\ref{sec:eq burning} and shown in the bottom panel.

\subsubsection{Relation to other recent schemes with broadened burning front}
\label{sec:others limiter}

Recently, \citet{Shen2018} used a burning limiter to broaden the burning front over a few cells which is different than the one used here and by  \citet{Kushnir2013}. In their implementation, changes in the temperature are limited within each hydrodynamical time-step, achieving broadened burning fronts and the ability to coverage consistently. There are, however, two problems with their approach. The first arises from the use of the global hydrodynamical time-step, $\Delta t$, instead of the specific cell sound crossing time, $t_s$. The time-step may be much shorter than the sound crossing time in a given cell, implying a lenient constraint, allowing significant changes within a cell sound crossing time. The second problem, is that they only limit the relative change of the temperature within $\Delta t$, which does not necessarily limit the energy changes. For example, in the $\rho_{0,7}=1$ case, the temperature changes by $4\%$ \citep[used by][as the maximal allowed change, in most runs]{Shen2018} between $x\sim2\,\textrm{cm}$ to $x\sim6\times10^{3}\,\textrm{cm}$, but in the same range $\mysim0.3\,\textrm{MeV}/m_p$ of thermonuclear energy is released, which increases the internal energy by more than $20\%$. In addition to the energy, a compositional change may occur without being captured by the temperature criteria. Compositional changes are also not taken into account by the limiter of \citet{Kushnir2013}, and are directly included in the modified version used here.  

\subsection{ASE burning}
\label{sec:eq burning}
 
The motivation for using the ASE scheme was discussed in Section~\ref{sec:demonstration}. In order to better explain the scheme, we begin with a brief description of the NSE and NSQE states of matter. 

\subsubsection{NSE and NSQE}
NSE is the unique nuclear composition of a system when strong and electromagnetic interactions are in a state of detailed balance for a given set of thermodynamic state variables and electron fraction. Applying detailed balance to the effective reaction that breaks up a nucleus with a nucleon number $A_{i}$ and a proton number $Z_{i}$ into free nucleons $(A_{i},Z_{i})\leftrightarrow Z_{i}p+N_{i}n$, where $N_{i}=A_{i}-Z_{i}$, yields a relation between the chemical potentials of the nucleus $\mu_{i}$, the free protons $\mu_{p}$ and the neutrons $\mu_{n}$: $\mu_{i}=Z_{i}\mu_{p}+N_{i}\mu_{n}$ \citep{Clifford1965}. The last relation can be written as
\begin{equation}\label{eq:NSE1}
Z_{i}\mu_{p}+N_{i}\mu_{n}=m_{i}c^{2}+k_{B}T\ln\left[\frac{n_{i}}{w_{i}(T)}\left(\frac{h^{2}}{2\upi m_{i}k_{B}T}\right)^{3/2}\right]+\mu_{i}^{\textrm{C}},
\end{equation}
where $m_i$ is the nuclear mass, $n_{i}$ is the number density, $w_i(T)$ is the nuclear partition function, $h$ is Planck's constant and $\mu_i^{\textrm{C}}$ is a Coulomb interaction term \citep{Calder2007,Seitenzahl2009}. The mass fractions of all nuclei in NSE can therefore be expressed in terms of the chemical potential of the protons and the neutrons and the nuclear binding energies $Q_{i}=(Z_{i}m_{p}+N_{i}m_{n}-m_{i})c^{2}$
 \begin{eqnarray}\label{eq:NSE2}
X_{i}&=&\frac{m_{i}}{\rho}w_{i}(T)\left(\frac{2\upi m_{i}kT}{h^{2}}\right)^{3/2} \nonumber\\
&\times&\exp\left[\frac{Z_{i}(\bar{\mu}_{p}+\mu_{p}^{\textrm{C}})+N_{i}\bar{\mu}_{n}-\mu_{i}^{\textrm{C}}+Q_{i}}{kT}\right],
\end{eqnarray}
where $\bar{\mu}_{p}=\mu_{p}-m_p c^2-\mu_p^C$ and $\bar{\mu}_{n}=\mu_{n}-m_n c^2$. Since the mass fractions of all nuclei must sum up to unity, $\sum_{i}X_{i}=1$, and the nuclear composition has the prescribed electron fraction, $Y_{e}\approx\sum_{i}X_{i}Z_{i}/A_{i}$, for a given $\rho$, $T$, and $Y_{e}$, the mass fractions of all the isotopes can be found by solving for $\bar{\mu}_{p}$ and $\bar{\mu}_{n}$ that satisfy the two constraints. The NSE state is found in this work by using a modified version of an NSE routine by Frank Timmes\footnote{http://cococubed.asu.edu/}. 

When nearing a state of NSE, the plasma may be in an intermediate state of NSQE \citep[][]{Bodansky1968}, in which neutrons, protons, and $\alpha$ particles are in a detailed balance, $\mu_{\alpha}=2\mu_{p}+2\mu_{n}$ and all heavier elements have a separate detailed balance with each other, such that the chemical potentials of every two of them, $i$ and $j$, satisfy $\mu_{i}-\mu_{j}=(N_{i}-N_{j})\mu_{n}+(Z_{i}-Z_{j})\mu_{p}$. In particular, the state of NSQE is uniquely determined by specifying $\rho$, $T$, $Y_{e}$, and $\tilde{Y}$ \citep[for a detailed discussion, see][]{Khokhlov89}. 

\subsubsection{Outline of ASE scheme}

The scheme we devise makes use of a generalization of NSQE. Assume that there are fast reactions that lead to equilibrium of $n$, $p$, and $\alpha$: $\mu_{\alpha}=2\mu_{p}+2\mu_{n}$. We can go over all reactions between all isotopes, and find the pairs for which the forward and the backward reactions are fast enough and that their values are close enough, such that the pair would reach an equilibrium during a cell sound crossing time (hereafter \textit{equilibrium reactions}). For example, assume we have a $(p,\gamma)$ reaction between isotope $A$ and isotope $B$, with abundances $Y_{A}$ and $Y_{B}$, respectively, such that this reaction has the following contribution to their abundances derivatives:
\begin{eqnarray}\label{eq:A and B}
\frac{dY_{A}}{dt}&=&-f_{lim}\langle\sigma v\rangle_f Y_{A}Y_{p}+f_{lim}\langle\sigma v\rangle_r Y_{B} \nonumber\\
\frac{dY_{B}}{dt}&=&f_{lim}\langle\sigma v\rangle_f Y_{A}Y_{p}-f_{lim}\langle\sigma v\rangle_r Y_{B}.
\end{eqnarray}
If the two positive terms, $b_f\equiv f_{lim}\langle\sigma v\rangle_f Y_{A}Y_{p}$ and $b_r\equiv f_{lim}\langle\sigma v\rangle_r Y_{B}$, are large enough
\begin{eqnarray}\label{eq:A and B fast}
\frac{Y_{A}}{\min(b_{f},b_{r})}&<&\epsilon t_s \nonumber\\
&\textrm{or}& \nonumber\\
\,\frac{Y_{B}}{\min(b_{f},b_{r})}&<&\epsilon t_s
\end{eqnarray}
where $\epsilon\ll1$, and their values are close enough
\begin{eqnarray}\label{eq:A and B close}
2\frac{|b_f-b_r|}{b_f+b_r}<\epsilon,
\end{eqnarray}
then within $t_s$ the two isotopes reach an equilibrium, in which $dY_{A}/dt=dY_{B}/dt=0$, such that
\begin{eqnarray}\label{eq:A and B NSE}
\frac{Y_{A}Y_{p}}{Y_{B}}=\frac{\langle\sigma v\rangle_r}{\langle\sigma v\rangle_f}=\left(\frac{Y_{A}Y_{p}}{Y_{B}}\right)_{\textrm{NSE}},
\end{eqnarray}
where the last equality is from detailed balance. Therefore, we can replace $Y_{A}$ and $Y_{B}$ with a single variable $\tilde{Y}_{AB}\equiv Y_{A}+Y_{B}$ in the rate equations. This saves one variable (and one reaction) and allows to derive $Y_{A}^{n+1}$ and $Y_{B}^{n+1}$ from $\tilde{Y}_{AB}^{n+1}$ by using Equation~\eqref{eq:A and B NSE}. Note that the requirement that the values of the forward and the backward reactions are close enough (Equation~\eqref{eq:A and B close}) is necessary, since a flow through a group of isotopes can shift their values from detailed balance to some other equilibrium state \citep[see detailed discussion in][]{Woosley1973}. 

\subsubsection{$n,p,\alpha$ equilibrium}

We make a few checks to decide whether neutrons, protons, and $\alpha$-particles are in equilibrium. The first check is to make sure that we are in the ballpark of such an equilibrium, by requiring that at the beginning of the burning step the deviation between the two ratios
\begin{eqnarray}\label{eq:npa check}
r&\equiv&\frac{Y_{\alpha}}{Y_n^2 Y_p^2},\nonumber\\
r_{\textrm{NSE}}&\equiv&\left(\frac{Y_{\alpha}}{Y_n^2 Y_p^2}\right)_{\textrm{NSE}},
\end{eqnarray}
is smaller than $50\%$. 

The second check is to find a fast reaction cycle that exchanges $\alpha$ with two $n$ and two $p$. An example for a cycle that can achieve such an exchange is \citep{Bodansky1968}
\begin{eqnarray}\label{eq:ag cycle}
^{32}\textrm{S}\left(\gamma,p\right)\left(\gamma,p\right)\left(\gamma,n\right)\left(\gamma,n\right)^{28}\textrm{Si}\left(\alpha,\gamma\right)^{32}\textrm{S}.
\end{eqnarray}
We look for cycles of equilibrium reactions that includes any of the following reactions (or their reverse):
\begin{enumerate} 
\item one $(\alpha,\gamma)$, two $(\gamma,p)$, two $(\gamma,n)$
\item one $(\alpha,p)$, one $(\gamma,p)$, two $(\gamma,n)$
\item one $(\alpha,n)$, two $(\gamma,p)$, one $(\gamma,n)$.
\end{enumerate} 
We demand that each step of the cycle is fast enough, ${Y}_{i}/\min(b_f,b_r)<\epsilon t_s$, with the relevant $b_f$ and $b_r$ and for all $i=n,p,\alpha$ that are participating in this step. Only if these two conditions are fulfilled, we proceed with the ASE scheme. Otherwise, we use the original burning step. Note that the application of ASE can change abruptly the compositions of $n$, $p$, or $\alpha$. This change can be controlled by the value of $\epsilon$, and it is quite small with the choices that we make for its value.

\subsubsection{Determination of the ASE configuration}

In order to find all equilibrium reactions and group the elements into effective abundances, we devise an algorithm that iteratively combines smaller groups into larger ones. At any point in this process, each isotope belongs to one group $q$ (possibly including only the isotope itself) with one of the groups including $n,p$, and $\alpha$ (and possibly other light elements, the light isotope group, LIG). Initially the LIG consists of exactly $n,p$, and $\alpha$ while all other isotopes are separated with each group having one isotope. 

At the beginning of each iteration, we calculate the abundance $\tilde{Y}_{q}$ of each group $q$,
\begin{eqnarray}\label{eq:variables}
\tilde{Y}_{q}=\sum_{l\in q}Y_{l}^{q},
\end{eqnarray}
where $Y_{l}^{q}$ is the abundance of the $l$-th isotopes inside the $q$-th group. Since each isotope $i$ is exactly in one group $q$, we can define $\tilde{Y}_{i}\equiv\tilde{Y}_{q}$. We then find among all reactions the fastest time-scale $t_{i,k}\equiv \tilde{Y}_{i}/\min(b_f(k),b_r(k))$ that satisfies the conditions of Equations~\eqref{eq:A and B fast} and \eqref{eq:A and B close}
\begin{eqnarray}\label{eq:conditions}
t_{i,k}&<&\epsilon t_s, \nonumber\\
2|b_f(k)-b_r(k)|/(b_f(k)+b_r(k))&<&\epsilon,
\end{eqnarray}
where $k$ is a reaction, $i$ is an isotope participating in $k$ (different from $n,p,\alpha$), and $b_f(k)$ and $b_r(k)$ are the forward and backward reaction rates of the reaction $k$  (similar to the definition below Equation~\eqref{eq:A and B}).  
We conservatively treat the following two simple cases: (1) There are exactly two isotopes in $k$ that are not in the LIG and $i$ is one of them. In this case we merge the groups of these two isotopes (if they are not already in the same group). (2) There is exactly one isotope in $k$ (not necessarily $i$) that is not in the LIG. In this case, we add this isotope to the LIG. Other more complicated cases (e.g. $^{12}$C$+^{16}$O$\leftrightarrow\alpha+^{24}$Mg) are not used for determining the ASE configuration. If no reaction is found that satisfies the conditions of Equation~\eqref{eq:conditions}, the iterations are terminated. 

The process of grouping described above is done once per hydrodynamic step (for each cell). Once it is finished, we are left with the task of calculating the evolution of the effective abundances of the groups by solving the appropriate ODEs throughout the duration of the hydro time-step. The ODEs consist of solving the rate equations for the abundances $\tilde{Y_q}$ of non-LIG groups (one abundance for each group), with $Q$ groups having more than one isotope and $J$ groups having one isotope (a total of $Q+J$ independent variables). The abundance of the LIG and the separate abundances of the different isotopes within each group are found by solving the algebraic relations of the equilibrium and conservation conditions. The rate of change of each non-LIG group $q$ is found by summing the reactions that involve isotopes in $q$
\begin{eqnarray}\label{eq:variables deriv}
\frac{d\tilde{Y}_{q}}{dt}=\sum_{l \in q}\frac{dY_{l}^{q}}{dt}.
\end{eqnarray}
Note that fast reactions that convert one isotope within a non-LIG group to another in the same group, such as $A(\alpha,\gamma)B$ do not affect $\tilde Y_q$ and are not included in the summation, thereby avoiding the resulting stiff equations \footnote{Reactions that only involves LIG isotopes are not included as well.}. The ODEs are solved using a semi-implicit Euler solver with adaptive time-steps. 
 
 If the grouping procedure at the beginning of the hydro time-step results in one group (NSE) there is no need for a solution of ODEs and all isotopes are found algebraically. More details about the scheme are provided in Appendix~\ref{sec:detailed}.

One caveat of the scheme described here, is that it depends on the value of the burning limiter $f_{lim}$. This can have a large effect for the FLASH simulations. For example, consider a cell very close to NSE conditions. Once a hydrodynamic step is made and before the burning iterations are preformed, the variables in the cell can temporarily (artificially) deviate significantly from NSE having large $|\dot{q}|$ and small $f_{lim}$. Since the ASE configuration is determined at the beginning of the burning step, it will use in this case a wrong value for $f_{lim}$. In order to resolve this issue we define new variables in each cell: $\dot{q}$, $\dot{Y}_{npa}$, and $\dot{\tilde{Y}}$ that are advected with the flow. This allows us to have a good estimate for $f_{lim}$ at the beginning of the burning step, with which we calculate the ASE configuration. In the case of V1D, the value of $f_{lim}$ from the previous time-step is used in order to calculate the ASE configuration. 

\subsubsection{Relation to previous schemes with generalized NSQE}

Previous attempts to devise a scheme that involves generalization of NSQE to more groups \citep{Hix2007,Parete-Koon2008b,Parete-Koon2008,Parete-Koon2010} pre-defined the isotope groups as a function of thermodynamic variables for some specific network. In this approach, the meaning of fast reactions is somewhat vague (for example, it does not depend on the sound crossing time of the cell) and the groups of the isotopes have to be calibrated manually for each case. In contrast, our scheme is adaptive and general.

\subsubsection{The implementation of the ASE burning in V1D}
\label{sec:new scheme V1D}

Following the force calculation, we decide whether the material is in ASE configuration or in NSE (with $\epsilon=0.01$). If the matter is not in any of these states, we proceed with the regular burning scheme described in Section~\ref{sec:old scheme V1D}. If the matter is in ASE configuration, we begin with a predictor step over $\Delta t$ with $\rho^n$ and $T^n$ to get the abundances, $\{Y_j\}^{n+1}$ and $\{\tilde{Y}_q\}^{n+1}$, at the end of the time-step. Since $\rho^{n+1}$ is known, we can use the ASE relations to find all $\{X_{i}\}(\rho^{n+1},T^{n+1},\{Y_j\}^{n+1},\{\tilde{Y}_q\}^{n+1})$ for a given $T^{n+1}$. For clarity, we write this as  $\{X_{i}\}(T^{n+1})$, and similarly $\varepsilon^{n+1}[\rho^{n+1},T^{n+1},\{X_i\}^{n+1}(T^{n+1})]$ and $p^{n+1}[\rho^{n+1},T^{n+1},\{X_i\}^{n+1}(T^{n+1})]$ will be written as  $\varepsilon^{n+1}(T^{n+1})$ and $p^{n+1}(T^{n+1})$, respectively. Then, we solve
\begin{eqnarray}\label{eq:GNSQE predictor}
&&\varepsilon^{n+1}\left(T^{n+1}\right)=\varepsilon^{n}- \\\nonumber 
&&\left(\frac{1}{\rho^{n+1}}-\frac{1}{\rho^n}\right)\left(\frac{p^{n}+p^{n+1}(T^{n+1})}{2}+q_{v}^n\right)+\Delta q\left[\{X_i\}^{n+1}\left(T^{n+1}\right)\right],
\end{eqnarray}
by iterating over $T^{n+1}$. A corrector step over $\Delta t$ follows with $\rho^{n+1}$, $T^{n+1}$, and $X_{i}^{n}$, to get a better estimate for the abundances, $\{Y_j\}^{n+1}$ and $\{\tilde{Y}_q\}^{n+1}$. We then re-solve Equation~\eqref{eq:GNSQE predictor}. It is important that the burning step is performed with the density and the temperature at the end of the time-step, because we want the equilibrium conditions to hold exactly at the end of the time-step. This is also the reason we use the density and the temperature at the end of the time-step for the regular burning scheme described in Section~\ref{sec:old scheme V1D}. Note that for a Lagrangian scheme, the same thermodynamic conditions and composition will be used for the next time-step, such that the equilibrium conditions are not altered between the end of one time-step and the beginning of the next. If the matter is in NSE, we can use the NSE relations to find all $\{X_{i}\}(\rho^{n+1},T^{n+1})$ for a given $T^{n+1}$, so we can solve Equation~\eqref{eq:GNSQE predictor} by iterating over $T^{n+1}$.

\subsubsection{The implementation of the ASE burning in FLASH}
\label{sec:new scheme FLASH}

We use $\rho^{n+1}$ and $\bar{T}^{n+1}=T(\rho^{n+1},\bar{\varepsilon}^{n+1},\bar{X}_{i}^{n})$ to decide whether the material is in ASE configuration or in NSE. For FLASH we are forced to use a less restrictive condition with $\epsilon=0.1$ (compared to $\epsilon=0.01$ in V1D), because of the composition advection that alter the condition $2|b_f-b_r|/(b_f+b_r)<\epsilon$. If the matter is neither in ASE configuration nor in NSE, then we proceed with the regular burning scheme described in Section~\ref{sec:old scheme FLASH}. If the matter is in ASE configuration, we calculate a burning step over $\Delta t$ with $\rho^{n+1}$ and $\bar{T}^{n+1}$ to get the abundances, $\{Y_j\}^{n+1}$ and $\{\tilde{Y}_q\}^{n+1}$, at the end of the time-step. Since $\rho^{n+1}$ is known, we can use the ASE relations to find all $\{X_{i}\}(\rho^{n+1},T^{n+1},\{Y_j\}^{n+1},\{\tilde{Y}_q\}^{n+1})$ for a given $T^{n+1}$. Then, we solve
\begin{eqnarray}\label{eq:GNSQE FLASH}
&&\varepsilon^{n+1}\left(T^{n+1}\right)=\bar{\varepsilon}^{n+1}+\Delta q\left[\{X_i\}^{n+1}\left(T^{n+1}\right)\right],
\end{eqnarray}
by iterating over $T^{n+1}$. If the matter is in NSE, we can use the NSE relations to find all $\{X_{i}\}(\rho^{n+1},T^{n+1})$ for a given $T^{n+1}$, and then re-solving Equation~\eqref{eq:GNSQE FLASH} by iterating over $T^{n+1}$.

\subsubsection{ASE performance}
\label{sec:ASE performance}

We now apply the ASE scheme for the V1D and FLASH simulations of the two test cases from Section~\ref{sec:limiter}. The density obtained with the two codes are shown at the lower panels of Figure~\ref{fig:New_scheme_CO_1e7} and Figure~\ref{fig:New_scheme_CO_1e8}. As before, the dashed blue vertical line marks the position where the burning limiter is no longer operating in the hydrocodes. The two dashed vertical black lines mark the positions where we first use the ASE scheme and the position where the material has reached NSQE state (the NSE state is not reached in these examples). The ASE burning changes the results for the $\rho_{0,7}=1$ very little, but performs faster ($\mysim2$ times faster for V1D and $\mysim6$ times faster for FLASH). For the $\rho_{0,7}=10$ case, the ASE scheme determines correctly the NSE position, and also significantly improves the performance ($\mysim21$ times faster for V1D and $\mysim56$ times faster for FLASH).

\subsection{A convergence study}
\label{sec:convergence}

A resolution of $\Delta x\sim\textrm{few}\times\textrm{km}$ is usually sufficient to resolve with a high accuracy the stellar scales for the densities considered in this work. The use of a burning limiter "sacrifices" $\mysim 1/f$ cells to correctly describe, at the resolved scales, the thermodynamic state and the compositions following a detonation wave. This would typically make the detonation wave widened over $\Delta x/f\sim\textrm{few}\times(10-100)\,\textrm{km}$, where smaller widening is obtained for larger $f$. Increasing $f$ also comes with the price of less accurate solution in the region where the limiter is operating, with an error of $\mysim f$. Therefore, we would like to use for $f$ as large a value as possible, as long as the solution within the region where the limiter is operating is reasonable. We can then choose $\Delta x$ to be small enough, such that $\Delta x/f$ is smaller than any relevant length-scale of the problem.

\begin{figure}
\includegraphics[width=0.48\textwidth]{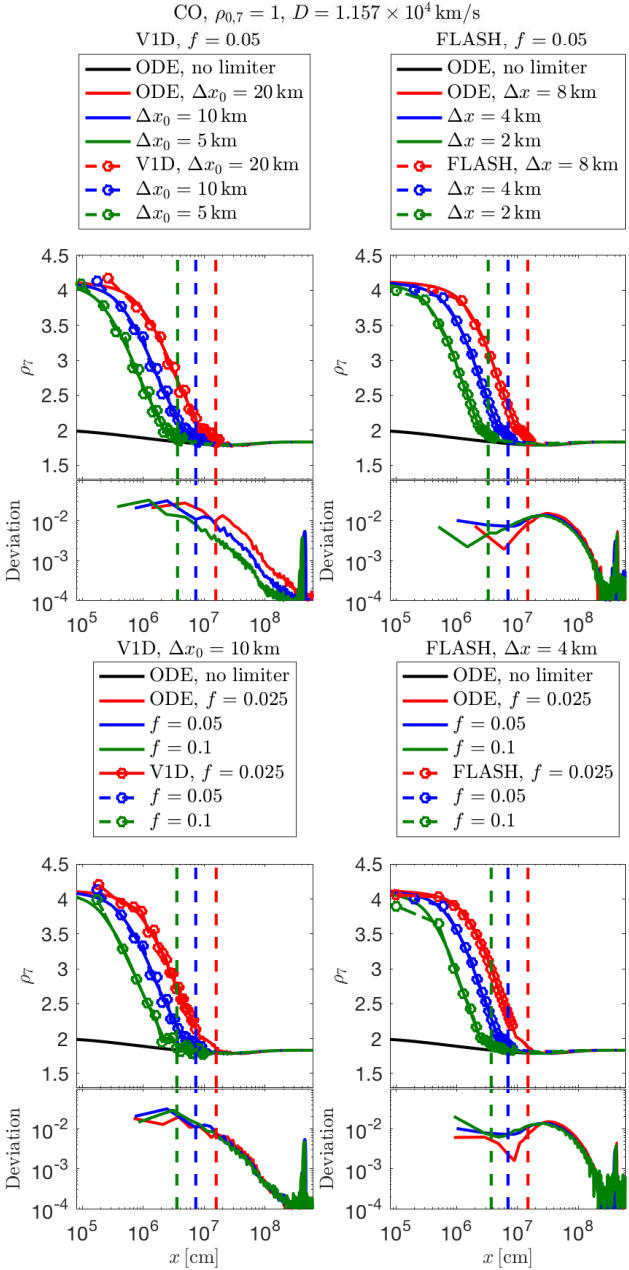}
\caption{The density profiles for a slightly overdriven detonation wave in CO with $\rho_{0,7}=1$ and $D=1.157\times10^{4}\,\textrm{km}\,\textrm{s}^{-1}$, as a function of the distance behind the shock. Top panels: the ODE solutions without a limiter (black) and with $f=0.05$ limiter are compared with the results obtained with V1D (left) and FLASH (right), using our new burning scheme with $f=0.05$, after the shock traversed a dynamical scale. We compare different resolutions, $\Delta x_{0}=20\,\textrm{km}$ (red), $\Delta x_{0}=10\,\textrm{km}$ (blue), and $\Delta x_{0}=5\,\textrm{km}$ (green) in the left-hand panel, and $\Delta x=8\,\textrm{km}$ (red), $\Delta x=4\,\textrm{km}$ (blue), and $\Delta x=2\,\textrm{km}$ (green) in the right-hand panel. The first 20 cells are marked with a circle and the dashed vertical lines mark the position where the burning limiter is no longer operating in the hydrocodes. Bottom panels: similarly to the top panels, but here we keep $\Delta x_{0}=10\,\textrm{km}$ and $\Delta x=4\,\textrm{km}$ fixed, while changing $f$. We compare $f=0.025$ (red), $f=0.05$ (blue), and $f=0.1$ (green). For clarity, we average the deviations of the numerical code solutions from the ODE solutions over five cells.
\label{fig:Convergence_CO_1e7}}
\end{figure}

\begin{figure}
\includegraphics[width=0.48\textwidth]{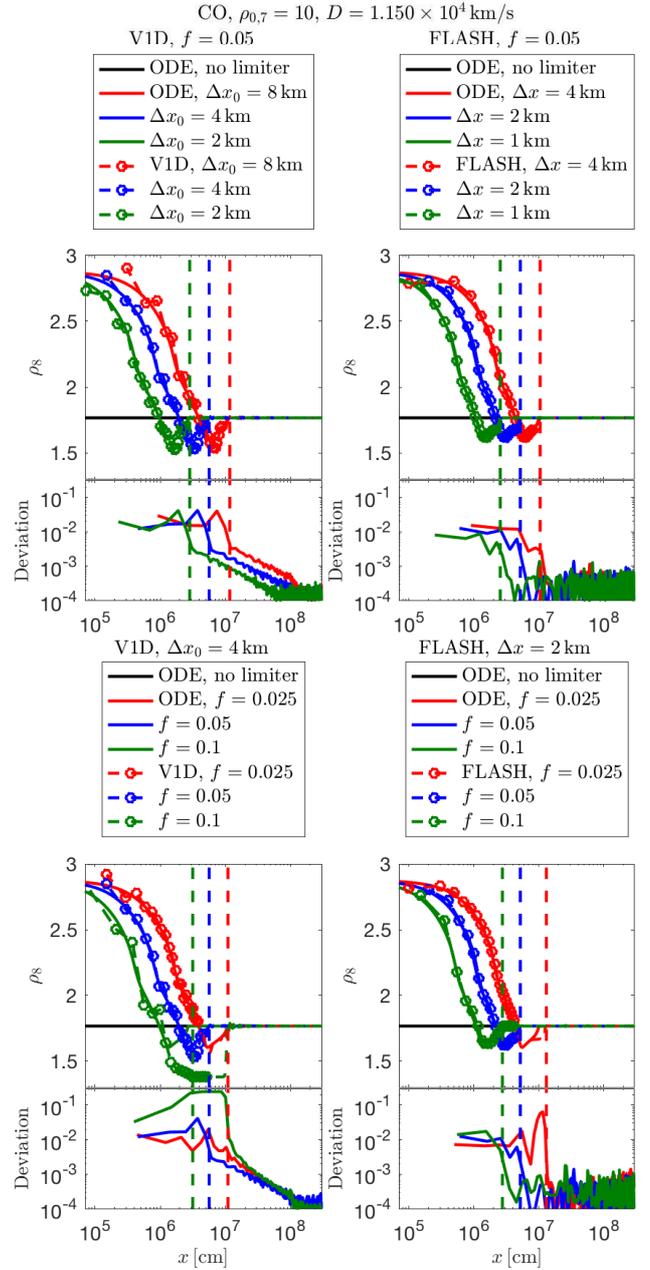}
\caption{Same as Figure~\ref{fig:Convergence_CO_1e7} for  $\rho_{0,7}=10$ and $D=1.150\times10^{4}\,\textrm{km}\,\textrm{s}^{-1}$. We compare in the top panels $\Delta x_{0}=8\,\textrm{km}$ (red), $\Delta x_{0}=4\,\textrm{km}$ (blue), and $\Delta x_{0}=2\,\textrm{km}$ (green) in the left-hand panel, and $\Delta x=4\,\textrm{km}$ (red), $\Delta x=2\,\textrm{km}$ (blue), and $\Delta x=1\,\textrm{km}$ (green) in the right-hand panel. In the bottom panels we use $\Delta x_{0}=4\,\textrm{km}$ and $\Delta x=2\,\textrm{km}$. 
\label{fig:Convergence_CO_1e8}}
\end{figure}

The densities obtained with the two codes, by using the new scheme, for the two test cases, where we changed both resolution and the value of $f$, are shown in Figure~\ref{fig:Convergence_CO_1e7} and Figure~\ref{fig:Convergence_CO_1e8}. The V1D results for the $\rho_{0,7}=1$ case (left-hand panels of Figure~\ref{fig:Convergence_CO_1e7}) follow the behaviour discussed above. The shock is widened over $\mysim(\Delta x_{0}/f)(\rho/\rho_0)$, and the error within the region where the limiter operates is $\mysim f$. Behind this region, the V1D solution quickly converges to the ODE solutions (which in turn, quickly converge to the ODE solution without limiter) to an accuracy of $\mysim10^{-3}$. In this case, it seems that $f=0.1$ is a good choice. The extended error at $x\sim4\times10^{8}\,\textrm{cm}$ is in fact related to inaccuracies of the ODE solver near NSE conditions (see detailed discussion in Section~\ref{sec:expansion}). At larger $x$ we just compare directly to the NSE condition at infinity, which can be calculated without solving the ODEs. The FLASH results for the $\rho_{0,7}=1$ case (right-hand panels of Figure~\ref{fig:Convergence_CO_1e7}) show similar behaviour to the V1D results, however there is a percent level error at $x\sim4\times10^{7}\,\textrm{cm}$ that is related to the less accurate boundary condition (see Section~\ref{sec:example results}). The V1D results for the $\rho_{0,7}=10$ case (left-hand panels Figure~\ref{fig:Convergence_CO_1e8}) are similar to the $\rho_{0,7}=1$, but in this case $f=0.1$ is too large and causes significant errors on scales in which the limiter is not operating. We therefore used $f=0.05$ as our default value. The FLASH results in this case (right-hand panels Figure~\ref{fig:Convergence_CO_1e8}) can be directly compared to the ODE solutions (the less accurate boundary condition has a negligible effect here), and they show the expected behaviour, as discussed above. The errors in the region where the limiter is not operating are smaller than $10^{-3}$. 

\subsection{Comparison of additional thermodynamic quantities and isotope abundances to the ODE solution} 
\label{sec:example results}

\begin{figure}
\includegraphics[width=0.48\textwidth]{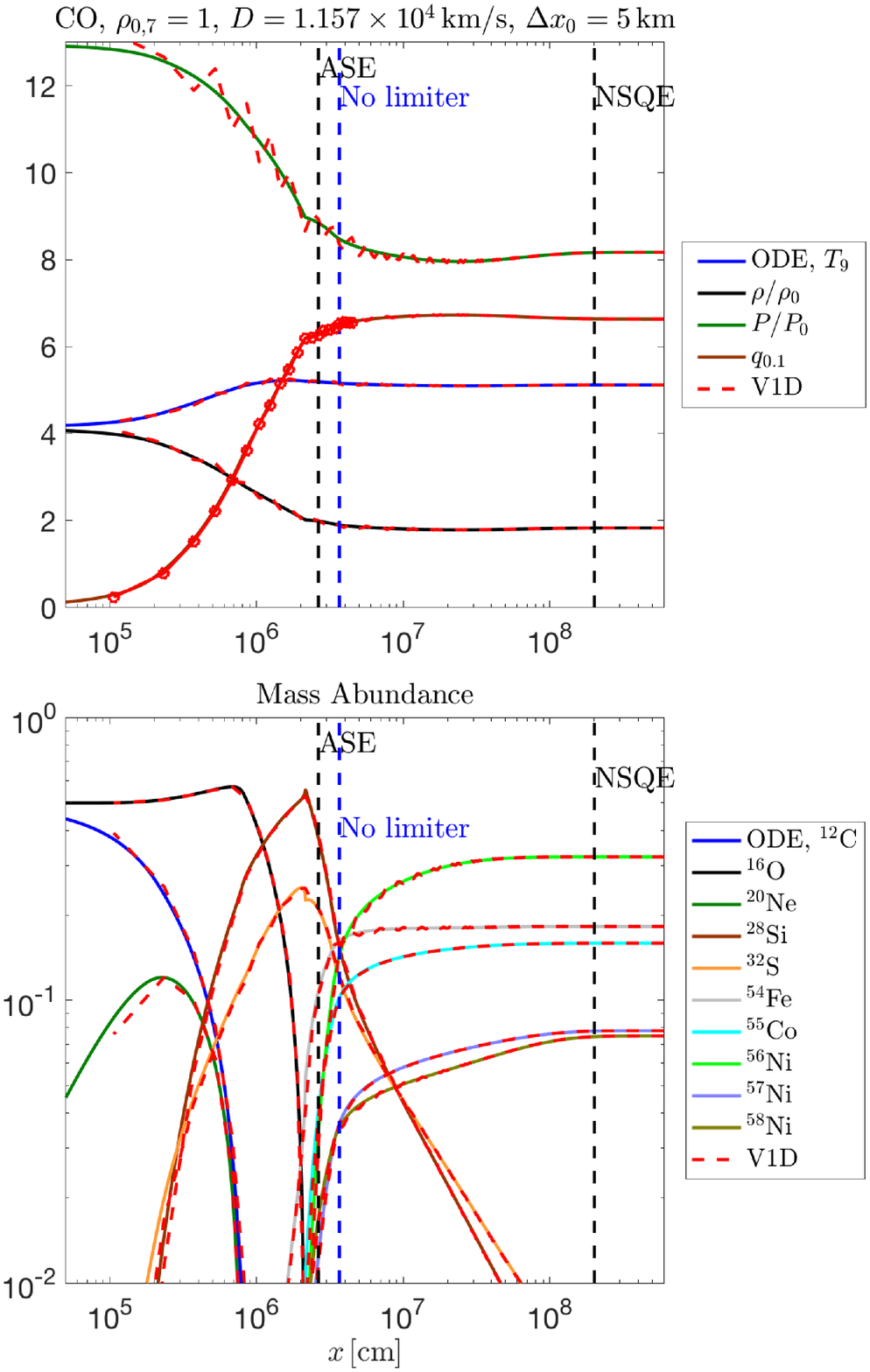}
\caption{
Calculation using V1D with the new scheme of the structure of a slightly overdriven detonation wave in CO with $\rho_{0,7}=1$ and $D=1.157\times10^{4}\,\textrm{km}\,\textrm{s}^{-1}$. Upper panel: profiles from the ODE solution, temperature (blue), density (black), pressure (green), and thermonuclear energy release (brown), are compared with the results obtained with V1D ($\Delta x_0=5\,\textrm{km}$, dashed red) after the shock traversed a dynamical scale. Bottom panel: Same as the upper panel for the mass fractions of a few key isotopes. The dashed vertical blue line marks the position where the burning limiter is no longer operating in V1D. The two dashed vertical black lines mark the positions where we first use the ASE scheme and the position where the material has reached NSQE state. 
\label{fig:MESA_V1D_Piston_CO_1e7}}
\end{figure}

In order to show that the new numerical solution accurately captures all aspects of the detonation wave, we compare in this section a few profiles of thermodynamic variables and mass abundances between the different solutions. In Figure~\ref{fig:MESA_V1D_Piston_CO_1e7}, we compare between the ODE solution and the V1D solution (with $\Delta x_{0}=5\,\textrm{km}$) of the $\rho_{0,7}=1$ case. All profiles behave similarly to the density profile discussed earlier. 

\begin{figure}
\includegraphics[width=0.48\textwidth]{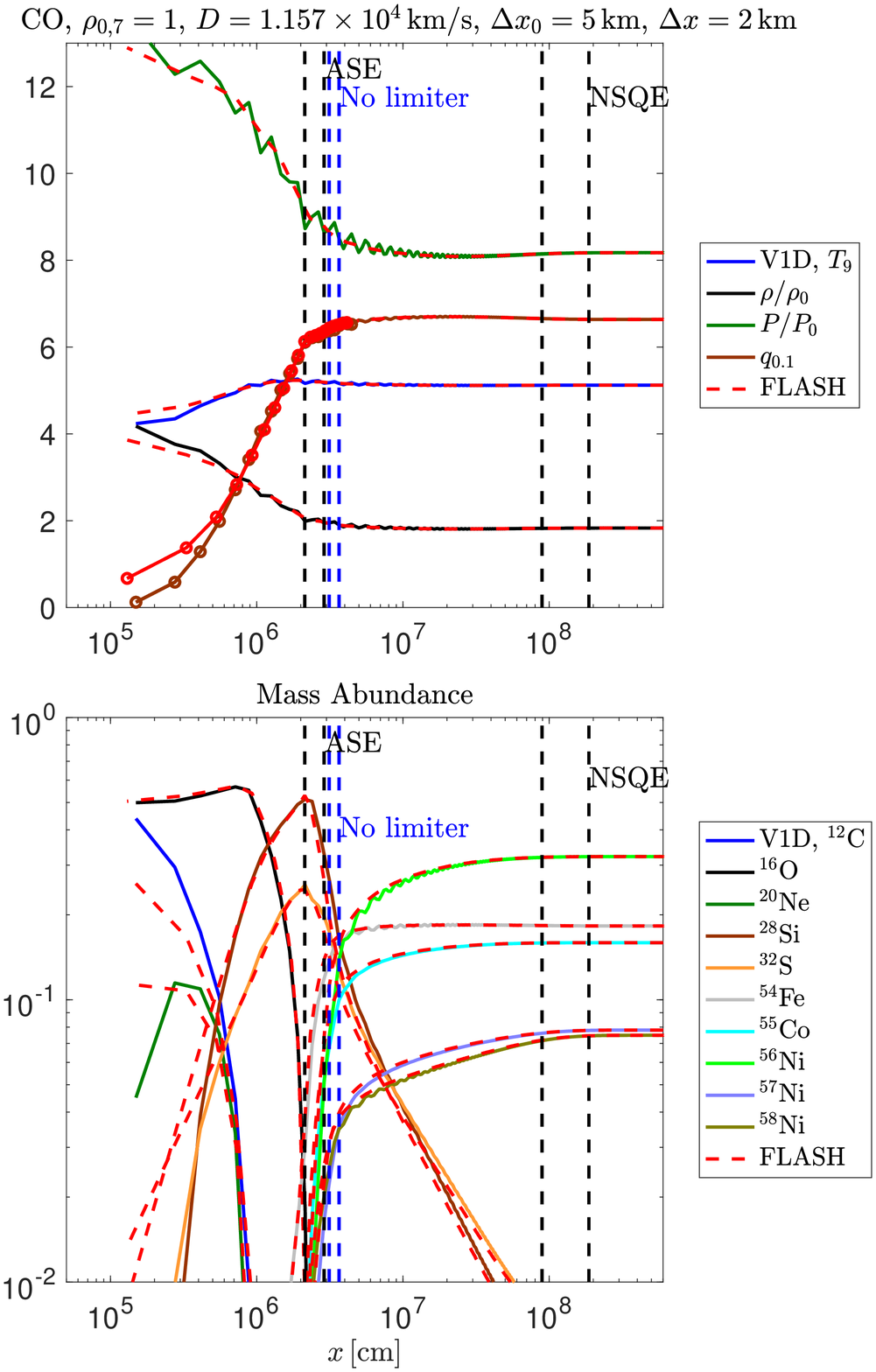}
\caption{Same as Figure~\ref{fig:MESA_V1D_Piston_CO_1e7}, except that solid lines are the results from V1D ($\Delta x_0=5\,\textrm{km}$ and zero piston velocity boundary condition) and red dashed lines are results from FLASH ($\Delta x=2\,\textrm{km}$). The dashed vertical lines are plotted for both hydrocodes. 
\label{fig:V1D_FLASH_Wall_CO_1e7}}
\end{figure}

We compare in Figure~\ref{fig:V1D_FLASH_Wall_CO_1e7} the FLASH results (with $\Delta x=2\,\textrm{km}$) to modified V1D calculations (with $\Delta x_0=5\,\textrm{km}$), which use the same simpler boundary condition of zero piston velocity. We note that in regions where the limiter is operating there should be a small difference between the two codes due to the different cell sizes, which cannot be matched exactly. Nevertheless, the deviation between the two codes is very small, which also demonstrate that both codes converge to the same solution. We note that there is still a small difference between the steady-state solution and the solution obtained with the zero piston velocity conditions (see the percent level error at $x\sim4\times10^{7}\,\textrm{cm}$ in the right-hand panels of Figure~\ref{fig:Convergence_CO_1e7}). This effect is more pronounced when the detonation wave does not reach a steady state before it crosses the dynamical scale, and so the boundary conditions still affect the structure of the detonation wave. While in the $\rho_{0,7}=1$ case, where the material approaches NSE roughly over the dynamical scale, the effect is small (and negligible for higher upstream densities), it is larger in lower upstream densities, see Section~\ref{sec:more cases}. 

\begin{figure}
\includegraphics[width=0.48\textwidth]{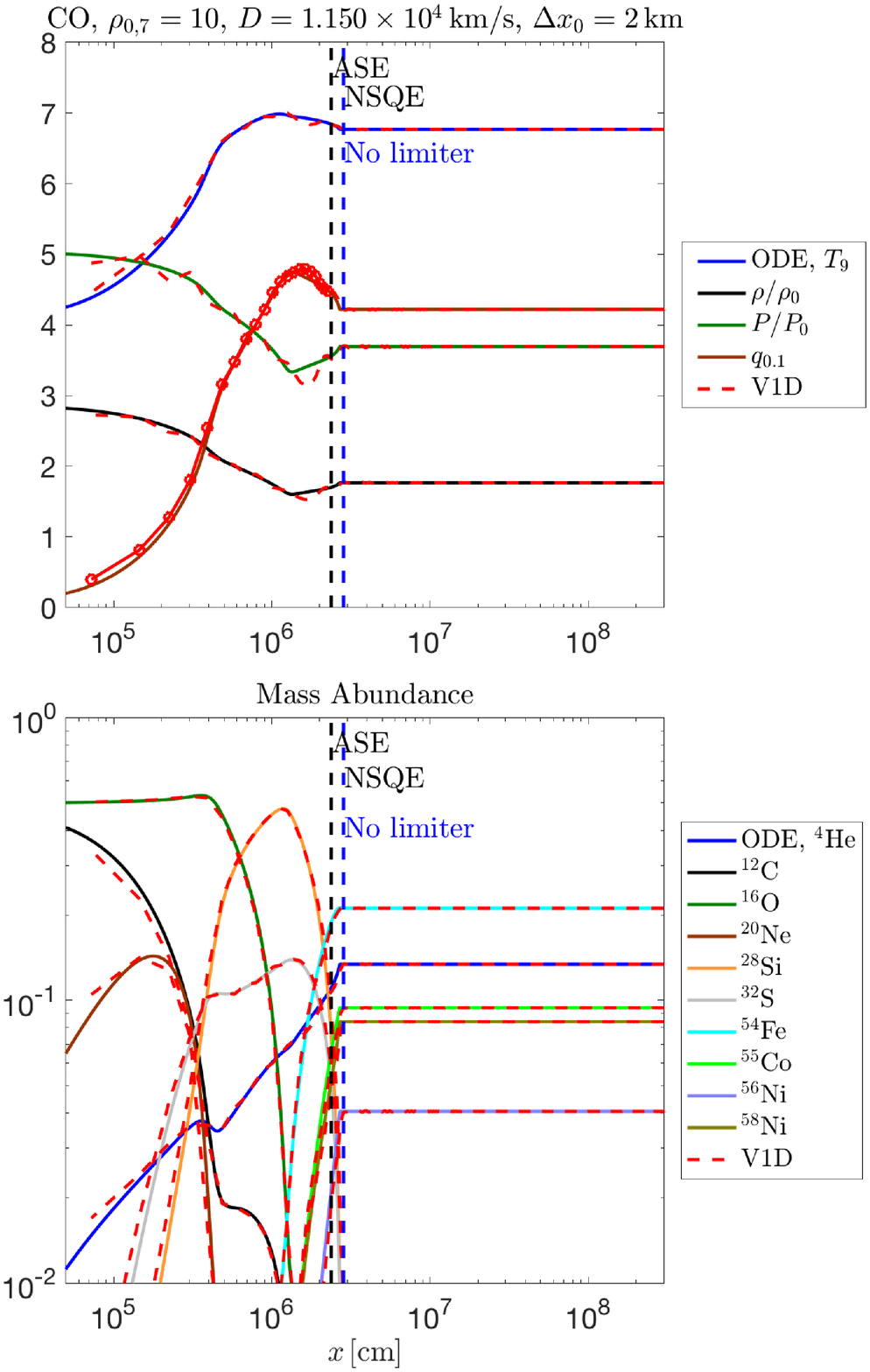}
\caption{Same as Figure~\ref{fig:MESA_V1D_Piston_CO_1e7}, for $\rho_{0,7}=10$, $D=1.150\times10^{4}\,\textrm{km}\,\textrm{s}^{-1}$ and $\Delta x_0=2\,\textrm{km}$. 
\label{fig:MESA_V1D_Piston_CO_1e8}}
\end{figure}

The case $\rho_{0,7}=10$ is simpler to compare since the steady solution is approached quickly (as NSE is approached over scales much smaller than the dynamical scale), and the boundary conditions have a small effect. The V1D results (with $\Delta x_{0}=2\,\textrm{km}$) are compared to the ODE solutions in Figure~\ref{fig:MESA_V1D_Piston_CO_1e8}. All profiles behave similarly to the density profile discussed earlier. Note that for all cases the reaction rates are slow enough to be resolved during a cell crossing time, such that a state of NSE is not reached. The FLASH results (with $\Delta x=1\,\textrm{km}$) behave similarly to the V1D results, and are not shown here.

\section{Results for more TNDW cases}
\label{sec:more cases}

In this section, we show the results for two more TNDW cases with low upstream density, $\rho_{0,7}=0.1$: CO (Section~\ref{sec:CO_1e6}) and He (Section~\ref{sec:He_1e6}), in order to demonstrate the robustness of the new scheme for different conditions. We obtained similar results for other upstream densities and initial compositions. 

\subsection{CO $\rho_{0,7}=0.1$}
\label{sec:CO_1e6}

\begin{figure}
\includegraphics[width=0.48\textwidth]{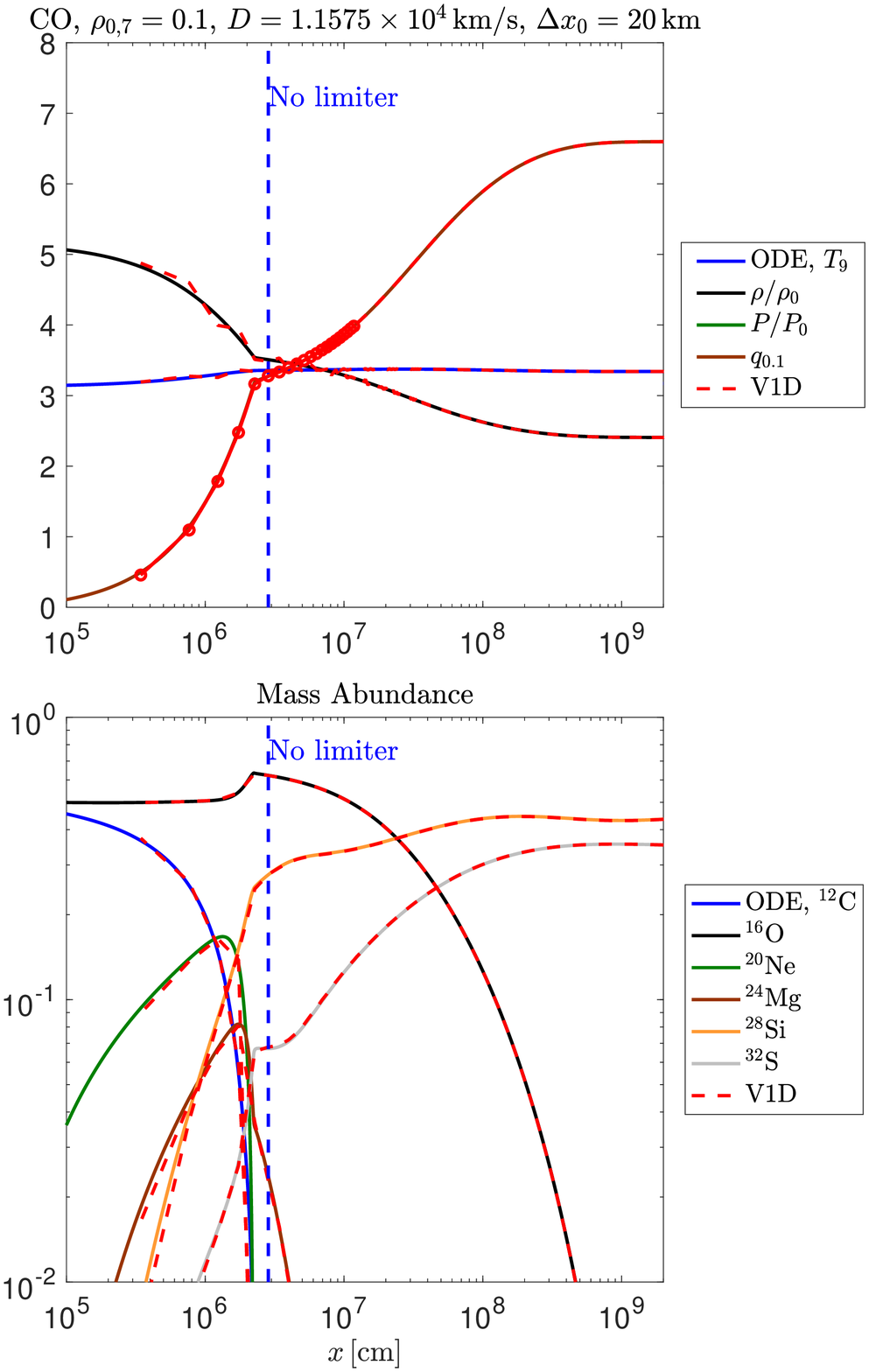}
\caption{Same as Figure~\ref{fig:MESA_V1D_Piston_CO_1e7}, for $\rho_{0,7}=0.1$, $D=1.1575\times10^{4}\,\textrm{km}\,\textrm{s}^{-1}$, and $\Delta x_0=20\,\textrm{km}$.
\label{fig:MESA_V1D_Piston_CO_1e6}}
\end{figure}

\begin{figure}
\includegraphics[width=0.48\textwidth]{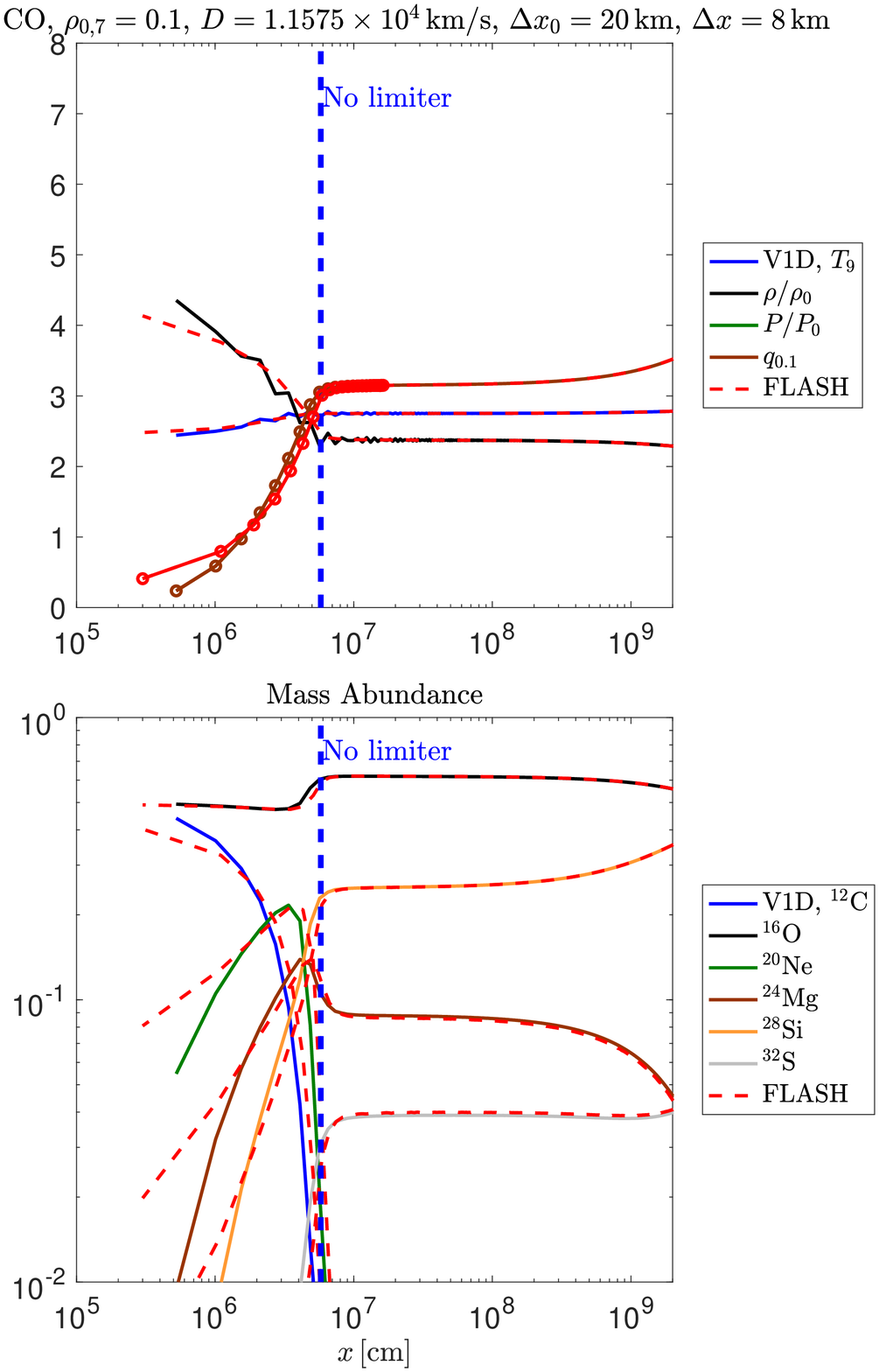}
\caption{Same as Figure~\ref{fig:MESA_V1D_Piston_CO_1e6}, except that solid lines are the results from V1D ($\Delta x_0=20\,\textrm{km}$ and zero piston velocity boundary condition) and red dashed lines are results from FLASH ($\Delta x=8\,\textrm{km}$). The dashed vertical blue lines are plotted for both hydrocodes.
\label{fig:V1D_FLASH_Wall_CO_1e6}}
\end{figure}

The relative difference between the pathological velocity and the CJ velocity for $\rho_{0,7}=0.1$ is smaller than $10^{-4}$, so we can choose  $D=1.1575\times10^{4}\,\textrm{km}\,\textrm{s}^{-1}$, which is slightly higher than the CJ velocity for this case, $D_{\textrm{CJ}}=1.1564\times10^{4}\,\textrm{km}\,\textrm{s}^{-1}$. We use initial cell sizes of $\Delta x_{0}=20\,\textrm{km}$ for V1D and  $\Delta x=8\,\textrm{km}$ for FLASH, such that a significant part of the energy release is not resolved, although the nuclear synthesis mostly is, see Figure~\ref{fig:CO_DetonationProp}. The approach to NSE in this case, is on scales much larger than the dynamical scale, so the zero piston velocity boundary condition has a large effect on the solution. In Figure~\ref{fig:MESA_V1D_Piston_CO_1e6}, we compare between the ODE solution and the V1D solution, with the more accurate boundary condition. In Figure~\ref{fig:V1D_FLASH_Wall_CO_1e6}, we compare the FLASH results to the V1D results, where both use the same boundary condition (zero piston velocity). Note that for this case, the reaction rates are slow enough, such that the ASE scheme is not used. The agreement between the different solutions is similar to the cases discussed in Section~\ref{sec:example results}. 

\subsection{He $\rho_{0,7}=0.1$}
\label{sec:He_1e6}

\begin{figure}
\includegraphics[width=0.48\textwidth]{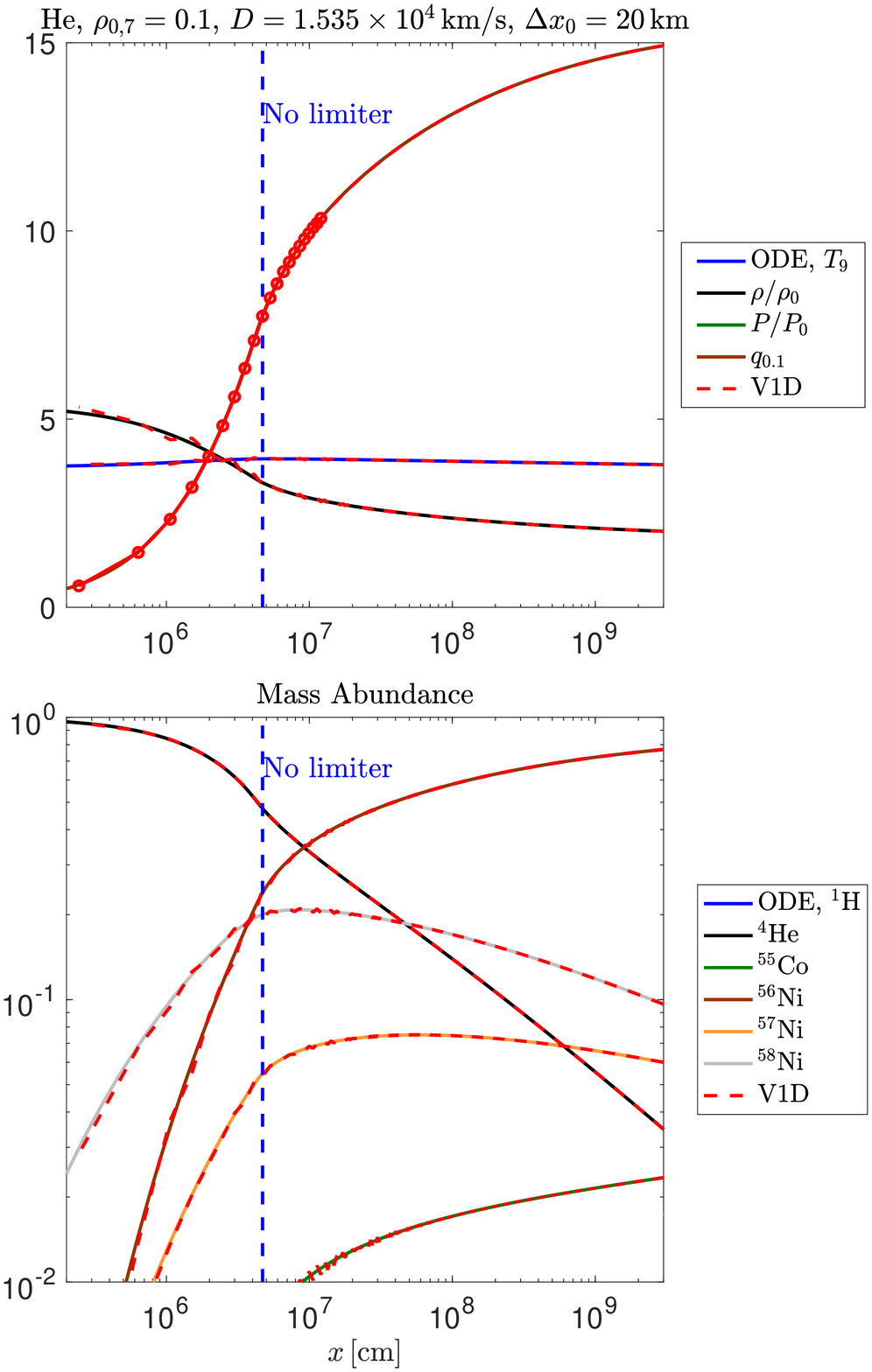}
\caption{Same as Figure~\ref{fig:MESA_V1D_Piston_CO_1e7}, for He, $\rho_{0,7}=0.1$, $D=1.535\times10^{4}\,\textrm{km}\,\textrm{s}^{-1}$, and $\Delta x_0=20\,\textrm{km}$..
\label{fig:MESA_V1D_Piston_He_1e6}}
\end{figure}

The structure of He detonations is quite different from the structure of CO detonations \citep{Khokhlov1984,Khokhlov1985,Kushnir2019}. We only mention here that the detonations are of the CJ type, where the energy release roughly follows the $^{4}$He depletion. For high upstream densities, $\rho_{0,7}\gtrsim0.015$, the burning of $^{4}$He synthesizes heavy elements with $\tilde{A}\equiv55$ much faster than the rate in which $^{4}$He is depleted. For the case studied here, $\rho_{0,7}=0.1$, the CJ velocity is $D_{\textrm{CJ}}=1.5342\times10^{4}\,\textrm{km}\,\textrm{s}^{-1}$, so we consider a slightly higher velocity, $D=1.535\times10^{4}\,\textrm{km}\,\textrm{s}^{-1}$, and we use an initial cell size of $\Delta x_{0}=20\,\textrm{km}$ for V1D. Here, a significant part of the energy release is not resolved, and $\tilde{A}=55$ is obtained on scales much smaller than the cell size. The approach to NSE is on scales much larger than the dynamical scale, so the zero piston velocity boundary condition has a large effect on the solution. In Figure~\ref{fig:MESA_V1D_Piston_He_1e6}, we compare between the ODE solution and the V1D solution, with the more accurate boundary condition. The agreement between the different solutions is similar to the cases discussed in Section~\ref{sec:example results}. Note that for this case, the reaction rates are slow enough, such that the ASE scheme is not used. We obtained a similar agreement (not shown here) between the V1D results and the FLASH results (with $\Delta x=8\,\textrm{km}$), where both use the same boundary condition (zero piston velocity).

\section{Results for an expansion problem}
\label{sec:expansion}

Following a TNDW in a supernova, the hot material expands and cools, which is usually accompanied with thermonuclear energy release and further nucleosynthesis. This phase can be difficult to calculate numerically, especially in cases where the material expands from an equilibrium configuration (e.g., NSE), such that there is a transition to a non-equilibrium configuration. The expansion phase provides an additional important test problem for numerical hydroburning schemes that are applied to global explosion simulations. The test problem consists of a homogenous region that is expanding in a prescribed manner that can be calculated exactly by solving appropriate ODEs for the (zero-dimensional) evolution in time. We will assume exponential expansion, with an e-folding time $\tau$, $\rho(t)\propto\exp(-t/\tau)$, such that 
\begin{equation}
\label{eq:expansion}
\frac{d\rho}{dt}=-\frac{\rho(t)}{\tau}.
\end{equation}
Finding the accurate (ODE) solution is challenging, since the numerical inaccuracy near NSE affects the ODE solver as well. We therefore construct in Section~\ref{sec:qnse solver}, a new ODE solver to accurately calculate the expansion problem. Then, in Section~\ref{sec:expansion case}, we test our new scheme against this solution for one particular case.

\subsection{A specialized ODE solver for the expansion problem}
\label{sec:qnse solver}

The challenging integration near NSE, discussed in Section~\ref{sec:TNDW challenge}, is also relevant for an ODE solver. For example, integration of Equations~\eqref{eq:ZND t} becomes inaccurate as the material approaches NSE. This was demonstrated by \citet{Kushnir2019} by monitoring the energy conservation of the solution, and in fact, for some cases it was not possible to integrate close enough to the NSE state with energy conservation better than $10^{-3}$. For expansion from NSE, the situation is much worse, since the initial condition causes significant inaccuracies. A common solution is to assume that NSE holds exactly if some (arbitrary) conditions hold, and solve the full ODE otherwise \citep{Iwamoto1999,Brachwitz2000,Lippuner2017}. For example, this approach was used for calculating the expansion problem in situations that are relevant for kilonova \citep{Lippuner2015}. Here, we provide an accurate ODE solver for the expansion problem.

The equation to solve for the expansion problem can be derived directly from energy conservation (we assume some given, constant $Y_e$)
\begin{eqnarray}\label{eq:expansion ODE}
&&d\varepsilon+PdV=dq\nonumber\\
&\Rightarrow&\frac{\partial\varepsilon}{\partial T}dT+\left(\frac{\partial\varepsilon}{\partial\rho}-\frac{P}{\rho^2}\right)d\rho+\sum_i\frac{\partial\varepsilon}{\partial X_i}dX_i=dq\nonumber\\
&\Rightarrow&\frac{dT}{dt}=\frac{1}{C_V}\left[\dot{q}-\sum_i\frac{\partial\varepsilon}{\partial X_i}\frac{dX_i}{dt}-\frac{d\rho}{dt}\left(\frac{\partial\varepsilon}{\partial\rho}-\frac{P}{\rho^2}\right)\right],
\end{eqnarray}
where $C_V$ is the heat capacity in constant volume and composition\footnote{Note that \citet{Shen2014} are missing a term in their equation (1) for the constant volume ($d\rho/dt=0$) case.}. Unless stated otherwise, the partial derivatives are taken with the rest of the independent variables remaining constant. Equations ~\eqref{eq:expansion} and~\eqref{eq:expansion ODE} together with Equations~\eqref{eq:fi} can be integrated from some NSE initial condition to calculate the expansion at any late time. However, direct integration of this set of equations would be highly inaccurate in this case, as discussed above. 

High numerical accuracy can be obtained by changing the variables of integration from $\{T,X_{i}\}$ to $\{T,\bar{X}_i\}$, where $\bar{X}_{i}=X_{i}-X_{0,i}(\rho,T)$ is the deviation of the mass fraction from its value in NSE, $X_{0,i}(\rho,T)$. With the new variables, Equation~\eqref{eq:expansion ODE} remains the same, and Equations~\eqref{eq:fi} are
\begin{eqnarray}\label{eq:fi exp}
\frac{d\bar{X}_{i}}{dt}=f_{i}(\rho,T,\{\bar{X}_{j}\})-\frac{\partial X_{0,i}}{\partial\rho}\frac{d\rho}{dt}-\frac{\partial X_{0,i}}{\partial T}\frac{dT}{dt}.
\end{eqnarray}
In this case, $f_{i}(\rho,T,\{\bar{X}_{j}\})$ can be calculated accurately arbitrarily close to the NSE state. For example, consider a $(p,\gamma)$ reaction between isotope $A$ and isotope $B$, with abundances $Y_{A}$ and $Y_{B}$, respectively. The contribution of this reaction to the $Y_A$ derivative can be written as
\begin{eqnarray}\label{eq:A gag}
\frac{dY_{A}}{dt}&=&-\langle\sigma v\rangle_f \left(Y_{0,A}+\bar{Y}_{A}\right)\left(Y_{0,p}+\bar{Y}_{p}\right)+\langle\sigma v\rangle_r \left(Y_{0,B}+\bar{Y}_{B}\right)\nonumber\\
&=&-\langle\sigma v\rangle_f \left(Y_{0,A}\bar{Y}_p+Y_{0,p}Y_{A}+\bar{Y}_{A}\bar{Y}_p\right)+\langle\sigma v\rangle_r\bar{Y}_{B},
\end{eqnarray}
where we defined $\bar{Y}_{i}=Y_{i}-Y_{0,i}(\rho,T)$ and we used
\begin{eqnarray}\label{eq:A0 and B0 NSE}
\frac{Y_{0,A}Y_{0,p}}{Y_{0,B}}=\frac{\langle\sigma v\rangle_r}{\langle\sigma v\rangle_f}.
\end{eqnarray}
In this form, the cancelation between two large terms is removed, and high numerical accuracy is maintained. The procedure described above can be generalized to any reaction, leading to the required set of ODEs. The calculation of the Jacobian terms is described in Appendix~\ref{sec:expansion jacobian}.

\subsection{An expansion from $\rho_7=20$, $T_{9}=7$, with $\tau=0.25\,\textrm{s}$}
\label{sec:expansion case}

\begin{figure}
\includegraphics[width=0.48\textwidth]{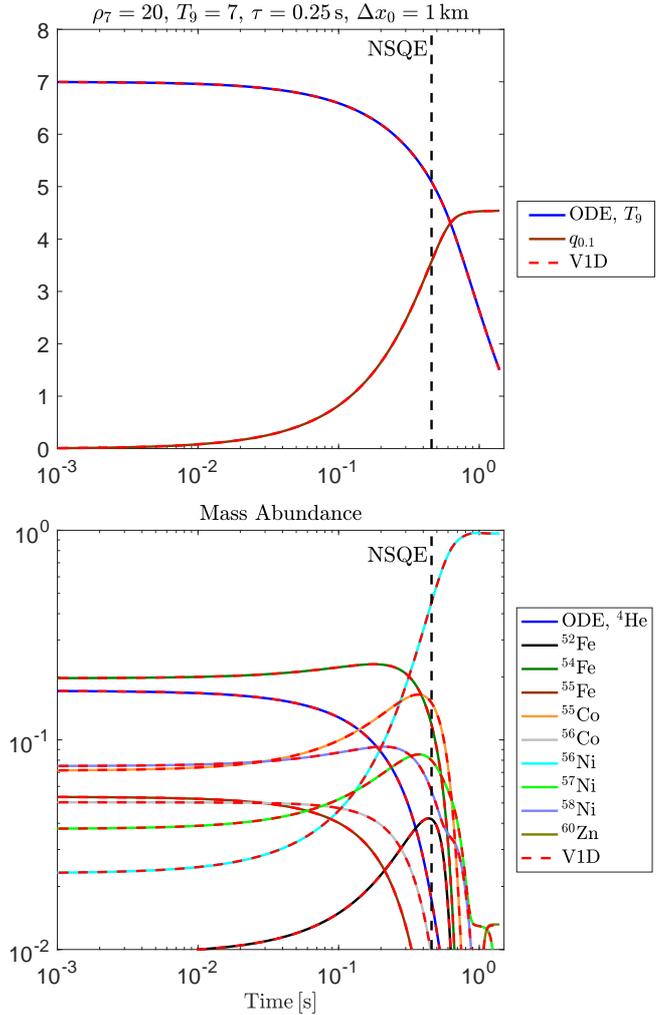}
\caption{Expansion from NSE with $\rho_{7}=20$, $T_{9}=7$, and $\tau=0.25\,\textrm{s}$ as a function of time. Upper panel: profiles from the ODE solution, temperature (blue), and thermonuclear energy release (brown) are compared with the results obtained with V1D ($\Delta x_0=1\,\textrm{km}$, dashed red). Bottom panel: Same as the upper panel for the mass fractions of a few key isotopes. In V1D, the reaction rates are slow enough to be resolved during a cell crossing time, such that a state of NSE is not reached in this example, and the initial state is NSQE. The vertical black lines mark the last time where the material was in NSQE state, followed directly by a calculation with the regular burning scheme (the ASE scheme was not triggered on these times). 
\label{fig:MESA_V1D_2e8_7e9_025_1km}}
\end{figure}

We now apply the ODE solver from the previous section to calculate the expansion from NSE with $\rho_{7}=20$, $T_{9}=7$ (roughly the NSE state of a CO detonation with $\rho_{0,7}=10$), and we choose $\tau=0.25\,\textrm{s}\sim1/\sqrt{G\rho}$. The result of the integration is shown in Figure~\ref{fig:MESA_V1D_2e8_7e9_025_1km}\footnote{We use \textsc{rodas4\_solver} and the parameters $\textsc{rtol}=10^{-7}$ (relative error tolerance) and $\textsc{atol}=10^{-10}$ (absolute error tolerance).}. As the material expands, it synthesizes $^{4}$He to heavier elements, releasing $\approx0.45\,\textrm{MeV}/m_p$. The composition at the end of the expansion, defined here at $T_9=1.5$, is dominated by $^{56}$Ni since $Y_e=0.5$. By using the new ODE solver from Section~\ref{sec:qnse solver}, integrating for $\{T,\bar{X}_{i}\}$, we were able to achieve energy conservation to better than $10^{-5}$. This is much better than the conservation obtained by integrating for $\{T,X_{i}\}$ which is of a few percents. The difference between the results, though, was on the sub-percent level.

Next we compare this solution to the results of the hydrocodes. In V1D, we calculate the spherical expansion of a single cell with $\Delta x_0=1\,\textrm{km}$. We use the same configuration as for the TNDW, expect that the inner boundary at $r=0$ is a solid wall and the outer boundary is an expanding piston with $v(t)=\Delta x_{0}/(3\tau)\exp(t/3\tau)$\footnote{Note that the mass within the cell is constant, so $\rho\propto r^{-3}$, where $r$ is the outer boundary of the cell. As a result, $\dot{\rho}=-3\rho v/r\Rightarrow v=r/(3\tau)\Rightarrow \dot{v}=v/(3\tau)$.}. The results of this run are plotted as dashed red lines in Figure~\ref{fig:MESA_V1D_2e8_7e9_025_1km}. Here, the reaction rates are slow enough to be resolved during a cell crossing time, such that a state of NSE is not reached in this example, and the initial state is NSQE. As the material expands, the regular burning scheme takes over. The agreement with the ODE is excellent. 

In FLASH, we calculated planar expansion with a fixed $\Delta x=1\,\textrm{km}$. Again, we use the same configuration as for the TNDW, expect that we impose $v(t)=\Delta x/\tau$ for the velocity of the (single) cell\footnote{Note that the volume of the cell is constant, so $\dot{\rho}=-\rho v/\Delta x\Rightarrow v=\Delta x/\tau$.}. We obtained similar results to the V1D case (not shown here).



\section{Summary and Discussion}
\label{sec:discussion}

We presented a new numerical scheme for thermonuclear burning (Sections~\ref{sec:demonstration} and~\ref{sec:new scheme}) that can be implemented in multidimensional full-star simulations. The new scheme allows an accurate calculation of TNDWs in a consistent way (i.e., without pre-describing the position and/or the conditions behind the TNDW) with all thermonuclear burning taking place in situ (without post-processing) for an arbitrary reaction network with hundreds of isotopes. We extensively test the new numerical scheme against a steady-state, planar TNDW (Sections~\ref{sec:new scheme} and~\ref{sec:more cases}). We showed that with resolutions that are typical for multidimensional full-star simulations, we can reach an accuracy better than a percent for the resolved scales, while keeping the numerical thermodynamic trajectory for unresolved scales within a few percent error from the true thermodynamic trajectory. We showed that at least in two cases, available codes do not reach robustly this level of accuracy, and may also perform inefficiently (by one to two orders of magnitude, without enforcing NSE under some arbitrary conditions). We further tested the new scheme for an expansion from NSE (Section~\ref{sec:expansion}), where we also presented a new accurate ODE solver for this problem. We provide subroutines, included in the online-only supporting information, that can be easily implemented in any hydrocode. These subroutines determine the ASE configuration and solve the associated ODEs.

While the demonstrations provided in this paper are restricted to steady-state TNDWs, where the thermodynamic trajectory is independent of reaction rates (Section~\ref{sec:limiter}), the burning limiter is applicable to the time-dependent TNDW expected in many supernova scenarios. As long as the length-scale on which the limiter operates ($\mysim\Delta x/f$) is much smaller than any other length-scale of the problem, then the assumption of a steady state in the region where the limiter operates is justified (since the solution in this region does not change while the region propagates a few times its own size), and we expect the converged solution to be exact. For example, consider the flow immediately following an ignition of a TNDW. The conditions for the ignition itself are usually obtained at low temperatures, such that the ignition region is resolved and the burning limiter is not triggered \citep[see detailed discussion and examples in][]{Kushnir2020}. As the TNDW forms, higher temperatures are obtained, and the burning limiter is operating in a non-steady-state region with a size comparable to the burning scale of a TNDW ($\mysim1\,\textrm{cm}$). Nevertheless, the TNDW quickly accelerates, and the steady-state assumption becomes valid for the region in which the burning limiter operates. We therefore expect that only a small region around the ignition location will be unresolved, such that any integral property of the simulation will rapidly converge to the correct value. Tests that show such a behaviour were conducted for a few cases, and will be reported in subsequent work. Our scheme is not directly applicable to scenarios that include length-scales that are much shorter than $\mysim\Delta x/f$, such as rapidly changing upstream conditions (e.g. the TNDW travels through density discontinuities, through a shockwave, or through a turbulent medium), and the operation of our scheme in these cases should be studied carefully.

Another example of a non-steady-state flow is an unstable TNDW. \citet{Khokhlov1993} showed that unsupported TNDWs propagating in CO with $\rho_{0,7}\gtrsim2$ are unstable with respect to longitudinal perturbations. This instability leads to non-linear oscillations of the detonation velocity and the following structure. However, this perturbation is damped at distances $x\gtrsim10\,\textrm{cm}$, such that it should play no role with km-scale cells, as indeed is the case in our simulations. Nevertheless, TNDWs propagating in CO were shown to be unstable to transverse perturbations as well \citep{Boisseau1996,Gamezo99,Timmes2000}, leading to the formation of a cellular detonation. These cells form on different scales, with some resolved by km-scale cells, while others too small to be resolved. It is not clear whether the formation of these small cells, which cannot be resolved in multidimensional full-star simulations, can affect the structure of TNDWs on larger scales, either by interaction with larger cells or by changing the average detonation velocity. The high effective resolutions obtained by our scheme are not applicable to this potential issue, and it remains an underlying assumption that these cells have no significant effect. 

A few technical caveats to the application of the new scheme in multidimensional full-star simulations should also be noted. The scheme contains two tuneable parameters ($f$ and $\epsilon$) that have to be calibrated for each set-up. We believe that our recommended values for these parameters should provide accurate results for a wide range of conditions, but each case should be tested carefully. A possible technical problem could be the use of a large number of isotopes in multidimensional simulations. While we showed that we can decrease the required time for integration of the rate equations, we did not provide a solution for the large memory requirement, as the new scheme assumes that the composition of all the isotopes is given at the beginning of the time-step. Addressing these challenges and applying the new scheme to global multidimensional supernova models are beyond the scope of this paper and will be done in future work. 

\section*{Acknowledgements}
We thank Eli Waxman for useful discussions. DK is supported by the Israel Atomic Energy Commission -- The Council for Higher Education -- Pazi Foundation and by a research grant from The Abramson Family Center for Young Scientists. BK is supported by the Beracha foundation and the Minerva Foundation with funding from the Federal German Ministry for Education and Research.









\begin{appendix}

\section{Input physics}
\label{sec:input}

Our input physics, which we briefly summarize below, are very similar to the ones used by \citet{Kushnir2019}, where more details can be found.

We use the NSE$5$ list of $179$ isotopes \citep{Kushnir2019} without $^{6}$He, such that our list is composed of $178$ isotopes. The nuclear masses were taken from the file \textsc{winvn\_v2.0.dat}, which is available through the JINA reaclib data base\footnote{http://jinaweb.org/reaclib/db/} \citep[JINA,][]{Cyburt2010}. For the partition functions, $w_{i}(T)$, we use the fit of \citet{Kushnir2019} for the values that are provided in the file \textsc{winvn\_v2.0.dat} over some specified temperature grid. 

The forward reaction rates are taken from JINA (the default library of 2017 October 20). All strong reactions that connect between isotopes from the list are included. Inverse reaction rates were determined according to a detailed balance. Enhancement of the reaction rates due to screening corrections is described at the end of this section. We further normalized all the channels of the $^{12}$C+$^{16}$O and $^{16}$O+$^{16}$O reactions such that the total cross-sections are identical to the ones provided by \citet{CF88}, while keeping the branching ratios provided by JINA. 

The EOS is composed of contributions from electron--positron plasma, radiation, ideal gas for the nuclei, ion--ion Coulomb corrections, and nuclear level excitations. We use the EOS provided by {\sc MESA} for the electron--positron plasma, for the ideal gas part of the nuclei, for the radiation and for the Coulomb corrections (but based on \citet{Chabrier1998} and not on \citet{Yakovlev1989}, see below). The electron--positron part is based on the \textit{Helmholtz} EOS \citep{Timmes00}, which is a table interpolation of the Helmholtz free energy as calculated by the Timmes EOS\footnote{http://cococubed.asu.edu/} \citep{Timmes1999} over a density--temperature grid with $20$ points per decade. This is different from \citet{Kushnir2019}, where the Timmes EOS was used for the electron--positron plasma, since the \textit{Helmholtz} EOS is more efficient and because the internal inconsistency of the \textit{Helmholtz} EOS \citep[see][for details]{Kushnir2019} is small enough within the regions of the parameter space studied here. We further include the nuclear level excitation energy of the ions, by using the $w_{i}(T)$ from above.

We assume that the Coulomb correction to the chemical potential of each ion is given by $\mu_{i}^{C}=k_{B}Tf(\Gamma_{i})$ and is independent of the other ions \citep[linear mixing rule (LMR),][]{Hansen1977}, where $k_{B}$ is Boltzmann's constant, $\Gamma_{i}=Z_{i}^{5/3}\Gamma_{e}$ is the ion coupling parameter, where $Z_i$ is the proton number, and $\Gamma_{e}\approx(4\upi\rho N_{A} Y_{e}/3)^{1/3}e^{2}/k_{B}T$ is the electron coupling parameter, where $N_{A}$ is Avogadro's number and $Y_e\approx\sum_i X_i Z_i/A_i$ is the electron fraction. We use the three-parameter fit of \citet{Chabrier1998} for $f(\Gamma)$.  Following \citet[][]{Khokhlov88}, we approximate the LMR correction to the EOS by $f(\Gamma)$ for a `mean' nucleus $\Gamma=\bar{Z}^{5/3}\Gamma_{e}$, where
\begin{eqnarray}
\bar{Z}=\frac{\sum_i Y_i Z_i}{\sum_i Y_i}.
\end{eqnarray}
The screening factor for a thermonuclear reaction with reactants $i=1,..,N$ and charges $Z_{i}$ is determined from detailed balance \citep{KushnirScreen}
\begin{eqnarray}\label{eq:NSE screening}
\exp\left(\frac{\sum_{i=1}^{N}\mu_{i}^{C}-\mu_{j}^{C}}{k_{B}T}\right),
\end{eqnarray}
where isotope $j$ has a charge $Z_{j}=\sum_{i=1}^{N}Z_{i}$ \citep[same as equation~(15) of][for the case of $N=2$]{Dewitt1973}.  

The input physics that we use in this paper are different from the ones used in \citet{Kushnir2019} in two aspects: we use a smaller list of $178$ isotopes \citep[most of the calculations in][were performed with the NSE$7$ list of $260$ isotopes]{Kushnir2019}, and we use the \textit{Helmholtz} EOS for the electron--positron instead of the more accurate Timmes EOS. Within the regions of the parameter space studied here, these differences have a negligible (deviation of less than $10^{-3}$) impact on the CJ and pathological detonation velocities (see Appendix~\ref{sec:structure}), and have a small (deviation of less than a percent) impact on the burning scales of the detonation waves.


\section{The structure of steady-state, planar TNDW}
\label{sec:structure}

In this appendix, we briefly summarize the known structure of a steady-state, planar TNDW, which can be solved accurately \citep{Imshennik1984,Khokhlov89,Townsley2016,Kushnir2019}. The structure of a detonation wave can be found by integration, where the initial conditions are the downstream values of the leading shock. We assume that the pressure, $P$, and the internal energy per unit mass, $\varepsilon$, are given as a function of the independent variables: density, $\rho$, temperature, $T$, and the mass fraction of the isotopes, $X_{i}$ ($\sum_{i}X_{i}=1$ and, unless stated otherwise, the sum goes over all isotopes). For planar, steady-state, non-relativistic hydrodynamics, the equations to integrate are \citep[see e.g.][]{Khokhlov89}
\begin{eqnarray}\label{eq:ZND}
d\rho&=&\frac{\frac{\partial P}{\partial T}\left(\frac{\partial\varepsilon}{\partial T}\right)^{-1}\left(dq-\sum_{i}\frac{\partial \varepsilon}{\partial X_{i}}dX_{i}\right)+\sum_{i}\frac{\partial P}{\partial X_{i}}dX_{i}}{u^{2}-c_{s}^{2}},\nonumber\\
dT&=&\left(\frac{\partial P}{\partial T}\right)^{-1}\left[\left(u^{2}-\frac{\partial P}{\partial\rho}\right)d\rho-\sum_{i}\frac{\partial P}{\partial X_{i}}dX_{i}\right],
\end{eqnarray}
where $c_{s}$ is the frozen (constant composition), non-relativistic speed of sound, $u$ is the velocity in the shock rest frame
\begin{eqnarray}\label{eq:u}
u=\frac{\rho_{0}}{\rho}D,
\end{eqnarray}
$\rho_{0}$ is the upstream density, $D$ is the shock velocity in the lab frame (in which the upstream fuel is at rest), $q$ is the average binding energy
\begin{eqnarray}\label{eq:q}
q=N_{A}\sum_{i}Q_{i}Y_{i},
\end{eqnarray}
$Q_{i}$ are the binding energies of the nuclei, $Y_{i}\approx X_{i}/A_{i}$ are the molar fractions of the nuclei, $A_{i}$ are the nucleon numbers, and $N_{A}$ is Avogadro's number. Unless stated otherwise, the partial derivatives are taken with the rest of the independent variables remaining constant. Upstream values will be denoted with subscript $0$. Equations~\eqref{eq:ZND}-\eqref{eq:q} are accurate as long as there is neither heat transfer nor particle exchange with the environment, which is an excellent approximation in our case \citep{Kushnir2019}. 

In order to calculate the structure of the detonation wave, a full derivative in time of Equations~\eqref{eq:ZND} is taken
\begin{eqnarray}\label{eq:ZND t}
\frac{d\rho}{dt}&=&\frac{\frac{\partial P}{\partial T}\left(\frac{\partial\varepsilon}{\partial T}\right)^{-1}\left(\frac{dq}{dt}-\sum_{i}\frac{\partial \varepsilon}{\partial X_{i}}\frac{dX_{i}}{dt}\right)+\sum_{i}\frac{\partial P}{\partial X_{i}}\frac{dX_{i}}{dt}}{u^{2}-c_{s}^{2}}\nonumber\\
&\equiv&\frac{\phi}{u^{2}-c_{s}^{2}},\nonumber\\
\frac{dT}{dt}&=&\left(\frac{\partial P}{\partial T}\right)^{-1}\left[\left(u^{2}-\frac{\partial P}{\partial\rho}\right)\frac{d\rho}{dt}-\sum_{i}\frac{\partial P}{\partial X_{i}}\frac{dX_{i}}{dt}\right].
\end{eqnarray}
The integration of Equations~\eqref{eq:ZND t} yields the state of a fluid element as a function of the time since it was shocked, given the reaction rates
\begin{eqnarray}\label{eq:fi}
dX_{i}/dt=f_{i}(\rho,T,\{X_{j}\}).
\end{eqnarray}
Equation~\eqref{eq:fi} includes the complexity of the problem, as many isotopes have to be included in the integration with many reactions. We present our results as a function of the distance behind the shock wave, $x$, connected to the time through $u=dx/dt$.

We briefly mention here the possible solutions of Equations~\eqref{eq:ZND t} \citep{Wood1960}. In the final state of the detonation wave all isotopes are in equilibrium, i.e. $dX_{i}/dt=0$ (for the case of a TNDW, this state is NSE, see Section~\ref{sec:eq burning}). The equilibrium composition is a function of the thermodynamic variables only, so there exists an equilibrium Hugoniot adiabat that connects to the upstream values. For a given shock velocity, $D$, the Rayleigh line that passes through the upstream values either does not intersect the equilibrium Hugoniot, is tangent to it (one point of intersection), or intersects it twice. The shock velocity for which there is one intersection is called the CJ velocity, and it is independent of reaction rates. If during the integration of Equations~\eqref{eq:ZND t} with $D=D_{\textrm{CJ}}$ the flow is always subsonic, then the minimal possible shock velocity is $D_{\textrm{CJ}}$. However, if during the integration the flow becomes sonic, then from Equations~\eqref{eq:ZND t}, we must require $\phi=0$ at the sonic point. The minimal shock velocity for which this condition is satisfied is called the pathological shock velocity, $D_{*}$, and it can only be found by integrating Equations~\eqref{eq:ZND t} (and so it depends on reaction rates). Overdriven detonations, which are solutions with higher shock velocities than the minimal shock velocity, either $D_{\textrm{CJ}}$ or $D_{*}$, exist as well, and they are subsonic throughout the integration. It can be shown that for pathological detonations, $\phi$ changes sign while crossing the sonic point, and so $d\rho/dt=0$ there. Finally, note that the equilibrium state is only approached asymptotically at an infinite distance behind the shock wave. 

\section{More details regarding the new burning scheme}
\label{sec:detailed}

We provide here some more details regarding the new burning scheme. 

\subsection{Determination of the ASE state}
\label{sec:GNSQE}

The objective is to find all abundances given $\rho$, $T$, $Y_{e}$, the abundances of the $J$ isotopes that are not grouped, $Y_{j}$ ($j\in J$), and the variables $\tilde{Y}_{q}$ ($q\in Q$). For a given guess of $\bar{\mu}_{p}$ and $\bar{\mu}_n$ we find all $\bar{Y}_{i}$ from the regular NSE relation (Equation~\eqref{eq:NSE2}). We use the given $Y_{j}$ instead of $\bar{Y}_{j}$ ($j\in J$) and we multiply each $\bar{Y}_{l}^{q}$ ($1\le l\le M(q)$, $q\in Q$) by $\tilde{Y}_{q}/\sum_{l=1}^{M(q)}\bar{Y}_{l}^{q}$. This completely define all abundances, so we can iterate on $\bar{\mu}_{p}$ and $\bar{\mu}_n$ until $\sum_{i}X_{i}=1$ and $\sum_{i}X_{i}Z_{i}/A_{i}=Y_{e}$. 

In order to calculate the Jacobian for the semi-implicit Euler solver, we need the derivatives $\partial Y_{k}/\partial Y_{j}$ and $\partial Y_{k}/\partial \tilde{Y}_{q}$ ($j\in J$, $q\in Q$, and $k\notin J$). To calculate these derivatives, we write $\bar{\mu}_{p}(\{Y_{j}\}_{j\in J},\{\tilde{Y}_{q}\}_{q\in Q})$ and $\bar{\mu}_{n}(\{Y_{j}\}_{j\in J},\{\tilde{Y}_{q}\}_{q\in Q})$ (for clarity we drop the dependence on $\rho$, $T$, and $Y_{e}$), such that each $Y_{k}$ is given as a function
\begin{eqnarray}\label{eq:Yk}
Y_{k}=&F_{k}&\left[\{\tilde{Y}_{q}\}_{q\in Q},\bar{\mu}_{p}\left(\{Y_{j}\}_{j\in J},\{\tilde{Y}_{q}\}_{q\in Q}\right),\right.\nonumber\\
&&\left.\bar{\mu}_{n}\left(\{Y_{j}\}_{j\in J},\{\tilde{Y}_{q}\}_{q\in Q}\right)\right],\,k\notin J.
\end{eqnarray}
In order to calculate $\partial Y_{k}/\partial Y_{j}$, we write
\begin{eqnarray}\label{eq:dYkdYj}
\frac{\partial Y_{k}}{\partial Y_{j}}=\frac{\partial F_{k}}{\partial \bar{\mu}_p}\frac{\partial \bar{\mu}_p}{\partial  Y_{j}}+\frac{\partial F_{k}}{\partial \bar{\mu}_n}\frac{\partial \bar{\mu}_n}{\partial  Y_{j}}.
\end{eqnarray}
Since $\partial F_{k}/\partial \bar{\mu}_p$ and $\partial F_{k}/\partial \bar{\mu}_n$ are given implicitly, we only need to calculate $\partial \bar{\mu}_p/\partial  Y_{j}$ and $\partial \bar{\mu}_n/\partial  Y_{j}$. To do this, we take full derivatives of the identities $\sum_{i}Y_{i}A_{i}=1$ and $\sum_{i}Y_{i}Z_{i}=Y_{e}$ to find
\begin{eqnarray}\label{eq:dYkdYj 2}
\frac{\partial \bar{\mu}_p}{\partial  Y_{j}}\sum_{k}\frac{\partial F_{k}}{\partial \bar{\mu}_p}A_{k}+\frac{\partial \bar{\mu}_n}{\partial  Y_{j}}\sum_{k}\frac{\partial F_{k}}{\partial \bar{\mu}_n}A_{k}=-A_{j},\nonumber\\
\frac{\partial \bar{\mu}_p}{\partial  Y_{j}}\sum_{k}\frac{\partial F_{k}}{\partial \bar{\mu}_p}Z_{k}+\frac{\partial \bar{\mu}_n}{\partial  Y_{j}}\sum_{k}\frac{\partial F_{k}}{\partial \bar{\mu}_n}Z_{k}=-Z_{j},
\end{eqnarray}
respectively. These are two equations for the two unknowns $\partial \bar{\mu}_p/\partial  Y_{j}$ and $\partial \bar{\mu}_n/\partial  Y_{j}$, which are easily solved. In order to calculate $\partial Y_{k}/\partial \tilde{Y}_{q}$, we write
\begin{eqnarray}\label{eq:dYkdYq}
\frac{\partial Y_{k}}{\partial \tilde{Y}_{q}}=\frac{\partial F_{k}}{\partial \tilde{Y}_{q}}+\frac{\partial F_{k}}{\partial \bar{\mu}_p}\frac{\partial \bar{\mu}_p}{\partial  \tilde{Y}_{q}}+\frac{\partial F_{k}}{\partial \bar{\mu}_n}\frac{\partial \bar{\mu}_n}{\partial  \tilde{Y}_{q}}.
\end{eqnarray}
Since $\partial F_{k/}\partial \tilde{Y}_{q}$ is given implicitly, we only need to calculate $\partial \bar{\mu}_p/\partial  \tilde{Y}_{q}$ and $\partial \bar{\mu}_n/\partial  \tilde{Y}_{q}$. As before, we take full derivatives of the identities $\sum_{i}Y_{i}A_{i}=1$ and $\sum_{i}Y_{i}Z_{i}=Y_{e}$ to find
\begin{eqnarray}\label{eq:dYkdYq 2}
\frac{\partial \bar{\mu}_p}{\partial  \tilde{Y}_{q}}\sum_{k}\frac{\partial F_{k}}{\partial \bar{\mu}_p}A_{k}+\frac{\partial \bar{\mu}_n}{\partial  \tilde{Y}_{q}}\sum_{k}\frac{\partial F_{k}}{\partial \bar{\mu}_n}A_{k}=-\sum_{l=1}^{M(q)}\frac{\partial F_{l}^{q}}{\partial \tilde{Y}_q}A_{l}^{q},\nonumber\\
\frac{\partial \bar{\mu}_p}{\partial  \tilde{Y}_{q}}\sum_{k}\frac{\partial F_{k}}{\partial \bar{\mu}_p}Z_{k}+\frac{\partial \bar{\mu}_n}{\partial  \tilde{Y}_{q}}\sum_{k}\frac{\partial F_{k}}{\partial \bar{\mu}_n}Z_{k}=-\sum_{l=1}^{M(q)}\frac{\partial F_{l}^{q}}{\partial \tilde{Y}_q}Z_{l}^{q},
\end{eqnarray}
respectively, where $F_{l}^{q}$ is the function that provides $Y_{l}^{q}$ (that has $A_{l}^{q}$ and $Z_{l}^{q}$). These are two equations for the two unknowns $\partial \bar{\mu}_p/\partial  \tilde{Y}_{q}$ and $\partial \bar{\mu}_n/\partial  \tilde{Y}_{q}$, which are easily solved. Note that
\begin{eqnarray}\label{eq:dYkdYq aux}
\sum_{l=1}^{M(q)}\frac{\partial F_{l}^{q}}{\partial \tilde{Y}_q}A_{l}^{q}=\sum_{l=1}^{M(q)}\frac{Y_{l}^{q}}{\tilde{Y}_q}A_{l}^{q}=\frac{1}{\tilde{Y}_q}\sum_{l=1}^{M(q)}Y_{l}^{q}A_{l}^{q},\nonumber\\
\sum_{l=1}^{M(q)}\frac{\partial F_{l}^{q}}{\partial \tilde{Y}_q}Z_{l}^{q}=\sum_{l=1}^{M(q)}\frac{Y_{l}^{q}}{\tilde{Y}_q}Z_{l}^{q}=\frac{1}{\tilde{Y}_q}\sum_{l=1}^{M(q)}Y_{l}^{q}Z_{l}^{q}.
\end{eqnarray}

\subsection{Calculation of the Jacobian}
\label{sec:Jacobian}

The reaction that are not ignored define $\dot{Y}_j$ ($j\in J$) and $\dot{Y}_l^{q}$ ($1\le l\le M(q)$, $q\in Q$), and then $\dot{\tilde{Y}}_q=\sum_{l=1}^{M(q)}\dot{Y}_l^{q}$. For the Jacobian we need to calculate four different types of terms. The first type is for $j1,j2\in J$
\begin{eqnarray}\label{eq:jacob 1}
\frac{d \dot{Y}_{j1}}{d Y_{j2}}=\sum_{l,q}\frac{\partial \dot{Y}_{j1}}{\partial Y_{l}^{q}}\frac{\partial Y_{l}^{q}}{\partial Y_{j2}}+\sum_{m=n,p,\alpha}\frac{\partial \dot{Y}_{j1}}{\partial Y_{m}}\frac{\partial Y_{m}}{\partial Y_{j2}}+\frac{\partial \dot{Y}_{j1}}{\partial Y_{j2}},
\end{eqnarray}
where all terms are given either by the rate routines or by the ASE routine (as explained in Appendix~\ref{sec:GNSQE}). Similarly, the second and the third type for $j\in J$ and $q1\in Q$ are given by
\begin{eqnarray}\label{eq:jacob 2}
\frac{d \dot{Y}_{j}}{d \tilde{Y}_{q1}}=\sum_{l,q}\frac{\partial \dot{Y}_{j}}{\partial Y_{l}^{q}}\frac{\partial Y_{l}^{q}}{\partial \tilde{Y}_{q1}}+\sum_{m=n,p,\alpha}\frac{\partial \dot{Y}_{j}}{\partial Y_{m}}\frac{\partial Y_{m}}{\partial \tilde{Y}_{q1}},
\end{eqnarray}
and by
\begin{eqnarray}\label{eq:jacob 3}
\frac{d \dot{\tilde{Y}}_{q1}}{d Y_{j}}&=&\sum_{l1=1}^{M(q1)}\frac{d \dot{Y}_{l1}^{q1}}{d Y_{j}}\\
&=&\sum_{l1=1}^{M(q1)}\left(\sum_{l,q}\frac{\partial \dot{Y}_{q1}}{\partial Y_{l}^{q}}\frac{\partial Y_{l}^{q}}{\partial Y_{j}}+\sum_{m=n,p,\alpha}\frac{\partial \dot{Y}_{q1}}{\partial Y_{m}}\frac{\partial Y_{m}}{\partial Y_{j}}+\frac{\partial \dot{Y}_{q1}}{\partial Y_{j}}\right),\nonumber
\end{eqnarray}
respectively. The last type for $q1,q2\in Q$ is given by
\begin{eqnarray}\label{eq:jacob 4}
\frac{d \dot{\tilde{Y}}_{q1}}{d \dot{\tilde{Y}}_{q2}}&=&\sum_{l1=1}^{M(q1)}\frac{d \dot{Y}_{l1}^{q1}}{d \dot{\tilde{Y}}_{q2}}\\
&=&\sum_{l1=1}^{M(q1)}\left(\sum_{l,q}\frac{\partial \dot{Y}_{q1}}{\partial Y_{l}^{q}}\frac{\partial Y_{l}^{q}}{\partial \tilde{Y}_{q2}}+\sum_{m=n,p,\alpha}\frac{\partial \dot{Y}_{q1}}{\partial Y_{m}}\frac{\partial Y_{m}}{\partial \tilde{Y}_{q2}}\right).\nonumber
\end{eqnarray}

\section{The Jacobian terms of the expansion ODE solver}
\label{sec:expansion jacobian}

For the Jacobian we need to calculate the derivatives of Equations~\eqref{eq:expansion ODE} and~\eqref{eq:fi exp} with respect to $\{T,\bar{X}_i\}$. The derivatives $\partial f_i/\partial T$ and $\partial f_i/\partial \bar{X}_j$ can be calculated directly from expressions in the form of Equation~\eqref{eq:A gag}, which would require the derivatives $\partial X_{0,i}/\partial T$ (given from the NSE relation, see below). Next, the relation
\begin{eqnarray}\label{eq:x xgag}
dX_{i}&=&\frac{\partial X_i}{\partial \bar{X}_i}d\bar{X}_i+\frac{\partial X_i}{\partial T}dT=d\bar{X}_i+\frac{\partial X_{0,i}}{\partial T}dT\nonumber\\
&\Rightarrow&\frac{\partial X_i}{\partial T}=\frac{\partial X_{0,i}}{\partial T},\;\;\;\frac{\partial X_i}{\partial  \bar{X}_i}=1,
\end{eqnarray}
can be used to calculate terms like
\begin{eqnarray}\label{eq:dCvdT}
\frac{\partial C_V}{\partial T}&=&\left(\frac{\partial C_V}{\partial T}\right)_{\{X_{i}\}}+\sum_i\left(\frac{\partial C_V}{\partial X_i}\right)_{T}\frac{\partial X_{0,i}}{\partial T},\nonumber\\
\frac{\partial C_V}{\partial \bar{X}_i}&=&\left(\frac{\partial C_V}{\partial X_i}\right)_{T},
\end{eqnarray}
where all terms are given either from the EOS or from the NSE relation. With similar terms one can calculate $\partial \dot{T}/\partial T$ and $\partial \dot{T}/\partial \bar{X}_i$. Next we have
\begin{eqnarray}\label{eq:dxidT}
\frac{\partial}{\partial T}\frac{d\bar{X}_i}{dt}&=&\frac{\partial}{\partial T}\left(f_i-\frac{\partial X_{0,i}}{\partial\rho}\dot{\rho}-\frac{\partial X_{0,i}}{\partial T}\dot{T}\right)\nonumber\\
&=&\frac{\partial f_i}{\partial T}-\frac{\partial ^2 X_{0,i}}{\partial \rho \partial T}-\frac{\partial ^2 X_{0,i}}{ \partial T^2}-\frac{\partial X_{0,i}}{\partial T}\frac{\partial\dot{T}}{\partial T},\nonumber\\
\frac{\partial}{\partial \bar{X}_j}\frac{d\bar{X}_i}{dt}&=&\frac{\partial}{\partial \bar{X}_j}\left(f_i-\frac{\partial X_{0,i}}{\partial\rho}\dot{\rho}-\frac{\partial X_{0,i}}{\partial T}\dot{T}\right)\nonumber\\
&=&\frac{\partial f_i}{\partial \bar{X}_j}-\frac{\partial X_{0,i}}{\partial T}\frac{\partial\dot{T}}{\partial \bar{X}_j},
\end{eqnarray}
which can be calculated with the NSE relation given below.

In order to finalize the calculation we need the following NSE derivatives, $\partial X_{0,i}/\partial T$,  $\partial X_{0,i}/\partial \rho$,  $\partial ^2 X_{0,i}/\partial T\partial \rho$, and $\partial ^2 X_{0,i}/\partial T^2$. To calculate these derivatives we write $\bar{\mu}_p(\rho,T)$ and $\bar{\mu}_n(\rho,T)$ (for clarity, we drop the dependence on $Y_e$), such that $X_{0,i}$ is given as a function
\begin{eqnarray}\label{eq:NSE deriv1}
X_{0,i}=F_{i}\left[\rho,T,\bar{\mu}_p\left(\rho,T\right),\bar{\mu}_n\left(\rho,T\right)\right].
\end{eqnarray}
Now we can evaluate
\begin{eqnarray}\label{eq:NSE deriv2}
\frac{d X_{0,i}}{d\rho}=\frac{\partial F_i}{\partial \rho}+\frac{\partial F_i}{\partial \bar{\mu}_p}\frac{\partial \bar{\mu}_p}{\partial \rho}+\frac{\partial F_i}{\partial \bar{\mu}_n}\frac{\partial \bar{\mu}_n}{\partial \rho}.
\end{eqnarray}
Since $\partial F_i/\partial \rho$, $\partial F_i/\partial \bar{\mu}_p$, and $\partial F_i/\partial \bar{\mu}_n$ are given implicitly, we only need to calculate $\partial \bar{\mu}_p/\partial \rho$ and $\partial \bar{\mu}_n/\partial \rho$. To do this, we take full derivatives of the identities $\sum_i X_{0,i}=1$ and $\sum_i X_{0,i}Z_i/A_i=1$ to find
\begin{eqnarray}\label{eq:NSE deriv3}
\frac{\partial \bar{\mu}_p}{\partial  \rho}\sum_{i}\frac{\partial F_{i}}{\partial \bar{\mu}_p}+\frac{\partial \bar{\mu}_n}{\partial  \rho}\sum_{i}\frac{\partial F_{i}}{\partial \bar{\mu}_n}=-\sum_i\frac{\partial F_{i}}{\partial\rho},\nonumber\\
\frac{\partial \bar{\mu}_p}{\partial  \rho}\sum_{i}\frac{Z_i}{A_i}\frac{\partial F_{i}}{\partial \bar{\mu}_p}+\frac{\partial \bar{\mu}_n}{\partial  \rho}\sum_{i}\frac{Z_i}{A_i}\frac{\partial F_{i}}{\partial \bar{\mu}_n}=-\sum_i\frac{Z_i}{A_i}\frac{\partial F_{i}}{\partial\rho},
\end{eqnarray}
respectively. These are two equations for the two unknowns  $\partial \bar{\mu}_p/\partial \rho$ and $\partial \bar{\mu}_n/\partial \rho$, which are easily solved. Similar line of arguments leads to equations that allows to calculate the rest of the derivatives. 

\end{appendix}

\bsp	
\label{lastpage}
\end{document}